\documentclass{aa}

\usepackage{amsmath}
\usepackage[table,xcdraw]{xcolor}
\usepackage{lscape}
\usepackage{capt-of}
\usepackage{graphicx}
\usepackage{placeins}
\usepackage[utf8]{inputenc}
\usepackage{filecontents}
\usepackage{txfonts}
\usepackage{natbib}
\newcommand{\mh}{[\rm M/\rm H]}
\newcommand{\Gaia}{{\it Gaia }}
\newcommand{\sm}{{\rm M}_\odot\, }

\begin{document}

\title{A box full of chocolates: The rich structure of the nearby stellar halo revealed by \Gaia and RAVE}

\titlerunning{Substructure in the Galactic stellar halo with \Gaia and RAVE}

\author{Amina Helmi\inst{1}  \and Jovan Veljanoski\inst{1} \and Maarten A. Breddels\inst{1} \and Hao Tian\inst{1}
      \and Laura V. Sales\inst{2}}

   \institute{Kapteyn Astronomical Institute, University of Groningen,
              Landleven 12, 9747 AD Groningen, The Netherlands\\
              \email{ahelmi@astro.rug.nl}
              \and
              Department of Physics \& Astronomy, University of California, Riverside, 900 University Avenue, Riverside,
               CA 92521, USA}

  \date{}


  \abstract
  {The hierarchical structure formation model predicts that stellar
    halos should form, at least partly, via mergers. If this was a
    predominant formation channel for the Milky Way's halo,
    imprints of this merger history in the form of moving groups or
    streams should exist also in the vicinity of the Sun.}
  {We study the kinematics of halo stars in the Solar neighbourhood
    using the very recent first data release from the \Gaia mission, and in
    particular the TGAS dataset, in combination with data from the RAVE
    survey. Our aim is to determine the amount of substructure present in the
    phase-space distribution of halo stars that could be linked to
    merger debris.  }
  {To characterise kinematic substructure, we measure the velocity
    correlation function in our sample of halo (low metallicity) stars. We
    also study the distribution of these stars in the space of energy
    and two components of the angular momentum, in what we call ``Integrals of Motion''
    space.}
  {The velocity correlation function reveals substructure in the form
    of an excess of pairs of stars with similar velocities, well above
    that expected for a smooth distribution. Comparison to
    cosmological simulations of the formation of stellar halos
    indicate that the levels found are consistent with the Galactic
    halo having been built fully via accretion. Similarly, the
    distribution of stars in the space of ``Integrals of motion'' is
    highly complex. A strikingly high fraction (between 58\% and upto more than 
    73\%) of the stars that are somewhat less bound than the Sun are on (highly) retrograde orbits.  A
    simple comparison to Milky Way-mass galaxies in cosmological
    hydrodynamical simulations suggests that less than 1\% have
    such prominently retrograde outer halos. We also identify several
    other statistically significant structures in ``Integrals of
    Motion'' space that could potentially be related to merger
    events.}{}


   \maketitle
%

\section{Introduction}

The hierarchical paradigm of structure formation predicts stellar
halos to be the most important repositories of merger debris
\citep{hw99}. Relics of accretion events may be in the form of
spatially coherent streams \citep{bj2005,cooper2010,helmi2011}, or in
moving groups of stars for events that happened a long time ago
\citep{hw99}. Data from wide area surveys such as SDSS and 2MASS has
revealed much spatial substructure and painted a picture in which the outer halo
(roughly beyond 20 kpc from the Galactic centre) was likely built purely via
mergers \citep{Newberg02,Majewski03,Belokurov2006a,Deason2015}. On the other
hand, the assembly process of the inner halo, where most of the halo stars
reside, is still unknown. Genuine halo streams crossing the Solar neighbourhood
have in fact been discovered nearly two decades ago \citep{h99}.  The
granularity of the nearby halo has been estimated by \citet{Gould2003}
who determined that streams, if present, should contain less than 5\%
of the stars each. These estimates are consistent with later
discoveries of substructures
\citep{Kepley2007,Morrison2009,Smith2009,Klement2010}.

Whether streams constitute just a minority of the halo is therefore
still under scrutiny. However, models predict that if fully built via
accretion, the stellar halo in the Solar neighbourhood should contain 300-500
streams originating mostly from a handful of massive progenitors
\citep{hw99,helmi2003,gomezhelmi2013}. Such a halo would appear
spatially very well-mixed and its granularity would only be truly
apparent in local samples with accurate kinematics with at least ten
times as many stars. Such large samples of stars with accurate full
phase-space information have yet to become available.

On the other hand, an important fraction of the inner halo may have
formed in-situ, either from a heated disk during merger events
\citep{Cooper2015}, or from the gas during early collapse
\citep{ELS,Zolotov2009}.  The idea of a ``dual'' nature of the stellar
halo has gained momentum from the kinematics and metallicities of
stars near the Sun \citep{Carollo2007, Carollo10}. A common explanation for this inner vs outer halo duality
are different formation paths, namely in-situ vs accretion origins
\citep{Tissera2014}.

The question of whether the halo was fully built by accretion or not
will most likely be answered with \Gaia data. The \Gaia mission was
launched by ESA in December 2013 and will collect data for a period of
at least 5 years. Its first data release, DR1, that has just become
available on  September 14, 2016, contains the positions on the sky and
G-magnitudes for over a billion stars. It also provides the proper motions
and parallaxes for over 2 million TYCHO-2 sources, in what is known as the
TYCHO-Gaia Astrometric Solution (TGAS) \citep{Gaia2016a,Lindegren2016}.

To make progress at this point in time on the accretion history of the
Milky Way, we have to partially rely on ground-based efforts to obtain
the required full phase-space information of halo stars. Several large
spectroscopic surveys have been carried out in the past decade, and
the one that matches TGAS best in terms of extent and magnitude
range  is RAVE \citep{Steinmetz2006}. The RAVE survey has
obtained spectra for more than 500,000 stars in the magnitude range
$9 < I < 12$. Its last data release, DR5 \citep{Kunder2016}, provides radial
velocities for over 400,000 independent stars, as well as
astrophysical parameters for a good fraction of these objects.

We take advantage of the powerful synergy between TGAS and RAVE to
construct a high quality dataset that allows us to explore the
formation and structure of the stellar halo. The results presented in
this paper may be considered as an appetizer of what can be expected,
as well as a teaser of the challenges to come when the next \Gaia data
release becomes available.

This paper is organised as follows. In Sec.~\ref{sec:data} we
introduce the dataset obtained by cross-matching TGAS to RAVE, from
which we construct a halo sample based on metallicity. In
Sec.~\ref{sec:results} we study the distribution of this sample of
halo stars in phase-space. To establish the presence of streams, we compute the velocity correlation function in
Sec.~\ref{sec:corr_f}, while in Sec.~\ref{sec:iom} we identify the
presence of several statistically significant substructures in the
space of ``Integrals of Motion'', i.e. defined by energy and two
components of the angular momentum. In Sec.~\ref{sec:disc} we discuss
our findings and make quantitative comparisons to cosmological
simulations of the formation of galaxies like the Milky Way.  We
present our conclusions in Sec.~\ref{sec:concl}.


\section{DATA}
\label{sec:data}

\subsection{TGAS and RAVE}
\label{sec:TGAS}

The \Gaia satellite has allowed a new determination of the astrometric
parameters (proper motions and parallaxes) for TYCHO-2 stars
\citep{Hoeg2000} by taking
advantage of the large time baseline between 1991.5 and 2015.0,
i.e. between the TYCHO epoch and the measurements obtained during
roughly the first year of science operations by the \Gaia mission
\citep{Lindegren2016}. As part of the \Gaia DR1 release the
astrometric parameters for $\sim 2\times 10^6$ stars (80\% of the
TYCHO-2 dataset, those with TGAS parallax error smaller than 1 mas), and
with magnitudes $ 6 \lesssim G < 13$ have been made publicly available to
the community \citep{Gaia2016a}. 
Most of these stars are within a few kpc from the Sun, while a few
objects exist at distances of $\sim 50$~kpc, such as supergiants in
the Large Magellanic Cloud.

The RAVE survey has obtained spectra for
$\sim 5 \times 10^5$ stars in the southern hemisphere since its start in 2003, from which
radial velocities have been derived \citep{Steinmetz2006}. If the
spectrum is of sufficient quality \citep[SNR $\ge$ 20,][]{Kordopatis2013}
then astrophysical parameters such as gravity, temperature and
metallicity, and hence absolute magnitude and distance can be
estimated.

We have made our own cross-match between TGAS and RAVE DR5, and found
210,263 stars in common. Although this is not a very large sample
(only 10\% of the full size of TGAS), it constitutes a very good
starting point to explore the dynamics and kinematics of stars near
the Sun. This is particularly true if we compare to what has been
possible with Hipparcos or TYCHO-2 in e.g. combination with the Geneva
Copenhagen Survey \citep{Nordstrom2004} or even with RAVE. Of this
cross-matched set, 203,992 stars have spectra with velocity error $\epsilon_{RV} \le
10$~km/s and {\sf CorrCoeff} $\ge 10$, indicating a good measurement
of the radial velocity. If we further consider only stars that satisfy
i) the SNR $\ge 20$ criterion and the flag {\sf algoConv} $\ne 1$ that
ensures a reliable determination of astrophysical parameters, and that
ii) have a relative parallax error $\le 30\%$ (be it from TGAS or
RAVE), the sample is reduced to 170,509 objects. For 29.5\% of the
stars in this set we use the RAVE parallaxes because they have a
smaller relative error than those in TGAS\footnote{We have checked
  that our results do not change significantly if we use the
  parallaxes with the smaller absolute error between TGAS and RAVE instead.}.

Although we have imposed a relative parallax error cut to have a good
quality dataset, the parallaxes of TGAS and RAVE could still be
subject to (different) systematic errors. In particular,
\citet{Binney2014} have performed a very careful comparison of the
parallaxes derived by the RAVE pipeline to those from other
methods. These included the parallaxes obtained by Hipparcos, the
distances to open clusters derived with giants or with dwarfs, and the
kinematic bias correction method of \citet{Sch2012}. In all cases,
\citet{Binney2014} found evidence that the parallaxes of RAVE giants
were underestimated by  $10 - 15\%$.

\begin{figure}
\centering
\includegraphics[height=5cm,clip=true]{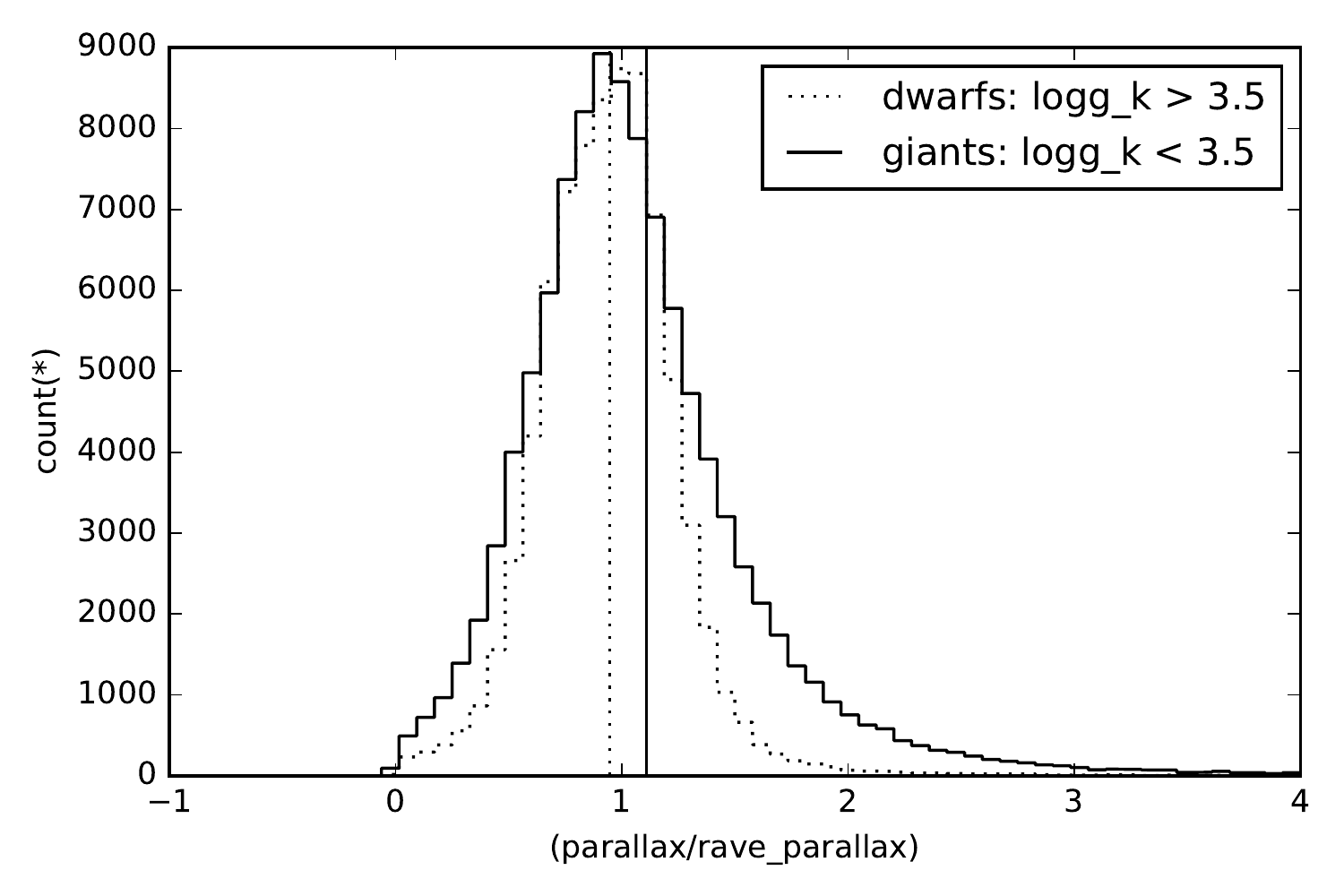} 
\caption{Distribution of the ratio between the parallax in TGAS to
  that in RAVE. The vertical lines indicate the mean ratio for dwarfs
  (dotted) and giants (solid) and shows that dwarfs' parallaxes are consistent with eachother in
  the datasets, but that the giants are
  systematically underestimated by 11\% in RAVE.}
\label{fig:par_comp}
\end{figure}

In Fig.~\ref{fig:par_comp} we plot the distribution of the ratio
between the TGAS and RAVE parallaxes for all stars satisfying the RAVE
quality criteria (SNR $\ge 20$ and {\sf algoConv} $\ne 1$) and with
TGAS positive parallax. The dotted histogram corresponds to dwarf
stars ($\log g \ge 3.5$) while the solid histogram to giants ($\log g
< 3.5$), and clearly shows the slight RAVE parallax underestimation in
the same sense and with similar amplitude as reported by
\citet{Binney2014}. We find that the amplitude of the bias is on
average 11\%, as indicated by the vertical solid line. We have
therefore decided to rescale the RAVE parallaxes for giant stars as
$\varpi_{RAVE}' = 1.11 \varpi_{RAVE}$, and in the rest of the paper we
work with this new parallax scale. We note here that the results
presented in this paper are not strongly dependent on this
scaling.

\begin{figure*}
\centering
\includegraphics[height=8cm,clip=true]{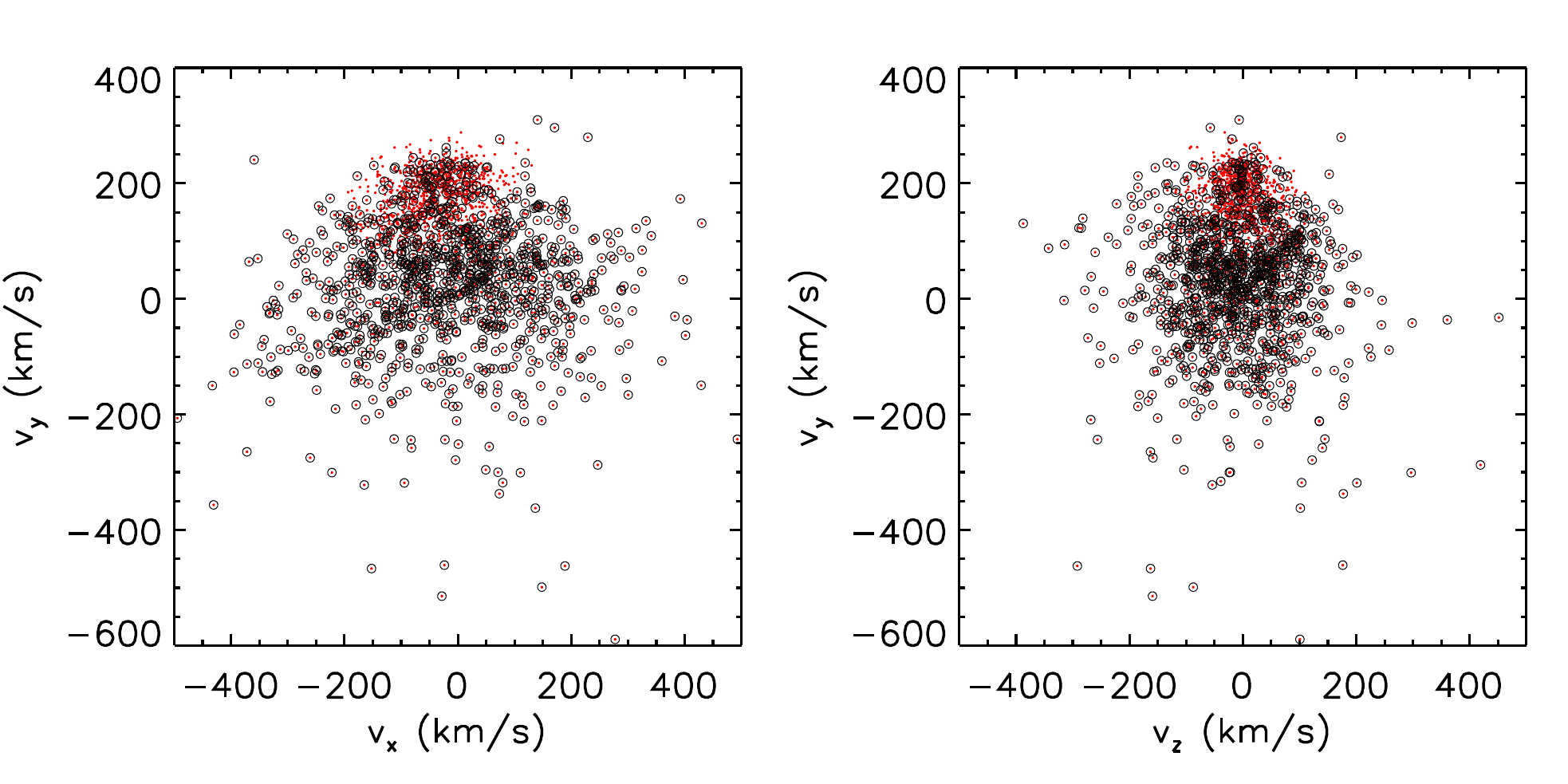} 
\caption{Velocities of stars in the TGAS cross-matched to RAVE sample
  described in Sec.~\ref{sec:TGAS} and \ref{sec:halo_sel}, selected
  according to the RAVE metallicity $\mh \le -1.5$ dex.  The subset of
  stars classified as belonging to the halo according to a
  two-component Gaussian kinematic model, are plotted with open
  circles.}
\label{fig:vel-feh}
\end{figure*}

\subsubsection{Coordinate transformations}

For the analysis presented below, we transform the
  coordinates measured for the stars ($\alpha, \delta, \varpi,
  \mu_\alpha, \mu_\delta, v_{los}$) into a Cartesian coordinate system.

  We compute the distance to a star by taking the reciprocal value of
  its parallax. Although this approach is not quite correct when the
  errors on the parallax are large
  \citep[see e.g.][]{Arenou99,Smith06,Astraatmadja16}, \citet{Binney2014}
  have shown that for RAVE stars the best distance indicator is
  1/$\langle \varpi \rangle$ where $\langle \varpi \rangle$ is the
  maximum likelihood estimate of the parallax given by the RAVE
  pipeline. Furthermore, in Appendix~\ref{App:distance-bias} we
  explore the effect of errors when the distance is calculated by
  inverting the parallax by using simulations based on the \Gaia
  Universe Model Snapshot \citep[GUMS,][]{Robin2012GUMS}. We find that
  the distance obtained in this way, even for relative parallax errors
  of order 30\%, coincides with the true distance, when assuming the
  parallax error distribution is Gaussian. This analysis, together
  with the fact that the TGAS and RAVE parallaxes are in good
  agreement, gives us confidence in the methodology used.

  We use the positions on the sky in Galactic $(l,b)$ coordinates,
  together with the distances to obtain the Cartesian $(x, y, z)$
  coordinates. To calculate the corresponding uncertainties, we used
  the errors on both $(l,b)$ and distance, as well as the covariance
  between $l$ and $b$. To compute the velocities, we first converted
  the proper motions from $(\mu_{\alpha}, \mu_{\delta})$ to $(\mu_{l},
  \mu_{b})$ as described in \citet{Poleski13}, where the $(\alpha$,
  $\delta)$ uncertainties and their covariance are all taken into
  account to obtain the errors on ($\mu_{l}, \mu_{b}$). Together with
  the line-of-sight velocity $v_{los}$, all these are used to derive
  $(v_{x}, v_y, v_z)$ using the transformations presented in
  \citet{Johnson87}. For the calculations of the associated
  uncertainties, we have propagated the errors in distance, $\mu_{l},
  \mu_{b}$, line-of-sight velocity, as well as the covariance between
  $\mu_{l}$ and $\mu_{b}$. Finally, we assume the Sun is located at
  8~kpc from the Galactic centre, and a Local Standard of Rest
  velocity $V_{LSR} = 220$~km/s in the direction of rotation (aligned
  with the $y$-axis at the location of the Sun). We do not correct for
  the peculiar motion of the Sun because for halo stars this will only
  introduce a negligible offset in the velocities.

\subsection{Construction of a halo sample}
\label{sec:halo_sel}

\subsubsection{Selection criteria}

The metallicities provided by the RAVE pipeline can be used to select
potential halo stars. In the set of 170,509 stars for which the
atmospheric parameters have been determined reliably and with relative
parallax errors smaller than 0.3, we find 2013 objects with $\mh \le
-1.5$, and with distances greater than 100~pc. The latter condition is used to
reduce the contamination by nearby disk stars.

We also consider another possibility, namely to select candidate
halo stars based on colours using the method developed by
\citet{Schlaufman2014}, also reported in \citet{Kunder2016}. The
combination of 2MASS and WISE colours selections $0.55 \le
(J_{2MASS}-K_{2MASS}) \le 0.85$ and $-0.04 \le$ (W1 - W2) $\le 0.04$
turns out to be very effective. This sample also still has some
amount of contamination by nearby disk red dwarfs, although
significantly reduced because of the WISE colour selection. Therefore we again
remove all stars within 100 pc from the Sun. This leaves us with a
sample of 1912 stars.

\subsubsection{Decomposition in disk and halo}

No selection will lead to a completely pure halo sample, although
Fig.~\ref{fig:vel-feh} shows that the level of contamination by (thin
and thick) disk stars is rather low for our RAVE metallicity selected
sample. Disk stars can be seen to cluster around $v_y \sim 200$~km/s,
while the halo stars on average have $v_y \sim 0$~km/s as the figure
clearly shows. Because we are interested in determining the level of
kinematic and phase-space substructure in this sample, and a disk may itself be considered as
a dominating (sub)structure, we proceed to flag
the stars by determining the probability that they belong to the
disk\footnote{We make no distinction between thin and thick disk in
  this analysis.} or the halo. To this end we have used the \textsc{sci-kit
  learn} package in \textsc{python} \citep{scikit-learn} and fit a two
component Gaussian Mixture Model to the Cartesian velocities of the
stars, $v_x$, $v_y$ and $v_z$. For this fit, we have considered the
full velocity covariance matrix, i.e. the resulting Gaussians'
principal axes are not necessarily aligned with the Cartesian
coordinate system. We find a very good fit, with one Gaussian centred
at $v_y \sim 180$~km/s that would model the disk
component(s)\footnote{Note that this mean velocity is lower than the
  LSR velocity assumed, and this could partly be because our low
  metallicity sample has contamination predominantly from the thick,
  and not the thin disk.}, while the second Gaussian is centred at
$v_y \sim 20$~km/s and would represent the likely halo stars. We then
flag stars to be members of the disk or halo components according to
whether they are more likely to belong to either of these two
Gaussians. Of the 2013 stars in our low metallicity sample which are plotted
in Fig.~\ref{fig:vel-feh}, we flag 1010 to belong to the disk, and 1003 to be likely halo stars (open circles).

This selection scheme however, creates a discontinuity in the
distribution of halo stars in velocity space, in the form of a hole at
the location of the disk.  To avoid any unwanted effects or spurious
results due to this hole, we additionally flag a number of stars,
previously marked to belong to the disk, as halo
candidates. Integrating the areas spanned by the 99\% confidence
isocontours of the two Gaussian components, we determine that
re-labelling 113 stars from the disk to the halo will effectively fill
the hole in the halo velocity distribution. We thus randomly draw 113
``disk'' stars following the distribution of the halo Gaussian, and
re-label them as ``halo'' stars. Our final halo sample has therefore
1116 stars. We have checked by creating 1000 sub-samples of 113
re-labelled disk-to-halo stars, that the results of the analysis we
are to present are robust.

We have also compared the distributions in velocity (and in other
projections of phase-space) of the RAVE metallicity selected sample to
that selected using the WISE colour criteria. We found the RAVE and
WISE samples to yield very similar distributions, with the WISE colour
selected sample being largely a subset of the RAVE metallicity selected
one. However, the metallicity selected sample has proportionally fewer
stars with disk-like kinematics and so we prefer to use it in the rest
of the paper.

\section{Analysis}
\label{sec:results}

Now that we have been able to compile a good sample of halo stars, we
can proceed to establish the amount of substructure present in the
Milky Way's stellar halo near the Sun. This will aid in
  understanding how important past accretion events were in the
  assembly of the Galaxy. On the other hand, the characterisation of
  the substructures found can tell us about the origin and nature of
  these potentially fundamental building blocks.

As discussed in the Introduction, we expect merger debris to be
apparent in velocity space in the form of tight moving groups of
stars, where a single progenitor galaxy could give rise to several of
these depending on its initial size \citep{hw99}. Therefore in
Sec.~\ref{sec:corr_f} we use a velocity correlation function to
establish their presence, which should reveal power on small scales
above that found in a smooth distribution.

We then turn to the space of ``Integrals of Motion'', which we define
in Sec.~\ref{sec:iom} by the stars' energy and two components of their
angular momenta.  If these were true integrals of motion\footnote{Note
  that energy is conserved if the gravitational potential is
  time-independent, while if the Milky Way were fully axisymmetric,
  only the $z$-component would be an integral of motion.}, we expect each accreted
object to define a clump whose extent depends on its initial size
\citep{hdz2000}. In reality, each of these clumps contains
substructure themselves \citep[corresponding to each of the groups
they produce in velocity space in a localised volume, see Fig.~5
of][]{gomez2010}, but given the current observational uncertainties
and the fact that the gravitational potential of the Milky Way is not
very well constrained, this (sub)substructure is in practise not yet
apparent.

Therefore, in Sec.~\ref{sec:retro} and \ref{sec:substr-iom} we search for the presence of clumps
in the space of ``Integrals of Motion''.  We establish the
significance of the various overdensities found by comparing to suitably
randomised smooth realisations of the data.

\subsection{Velocity correlation function}
\label{sec:corr_f}

To quantify the presence of moving groups or streams in our
dataset, we compute the velocity correlation function defined as
\begin{equation}
\xi({\bf \Delta v}) = \frac{\langle DD \rangle}{\langle RR \rangle} - 1
\label{eq:xi}
\end{equation}
where $\langle DD \rangle$ is the number of pairs in the data that have a
velocity difference $|{\bf v}_i - {\bf v}_j| = \Delta$ in a given
range, $\Delta_k, \Delta_{k+1}$, and similarly $\langle RR \rangle$
corresponds to the number of random pairs. The velocity correlation
function therefore can be used to establish if the data depicts a
statistically significant excess of pairs compared to random sets
which would then indicate the presence of velocity clumping or streams.

For this statistical test we have generated 1000 random samples whose
velocity distributions follow the observed 1D distributions, but where
the velocity components have been reshuffled.  This ensures that we break
any correlations or small scale structure present in the data in a
model independent way. In this particular case, we have scrambled the
$v_y$ and $v_z$ velocities.

\begin{figure}
\centering
\includegraphics[width=9cm]{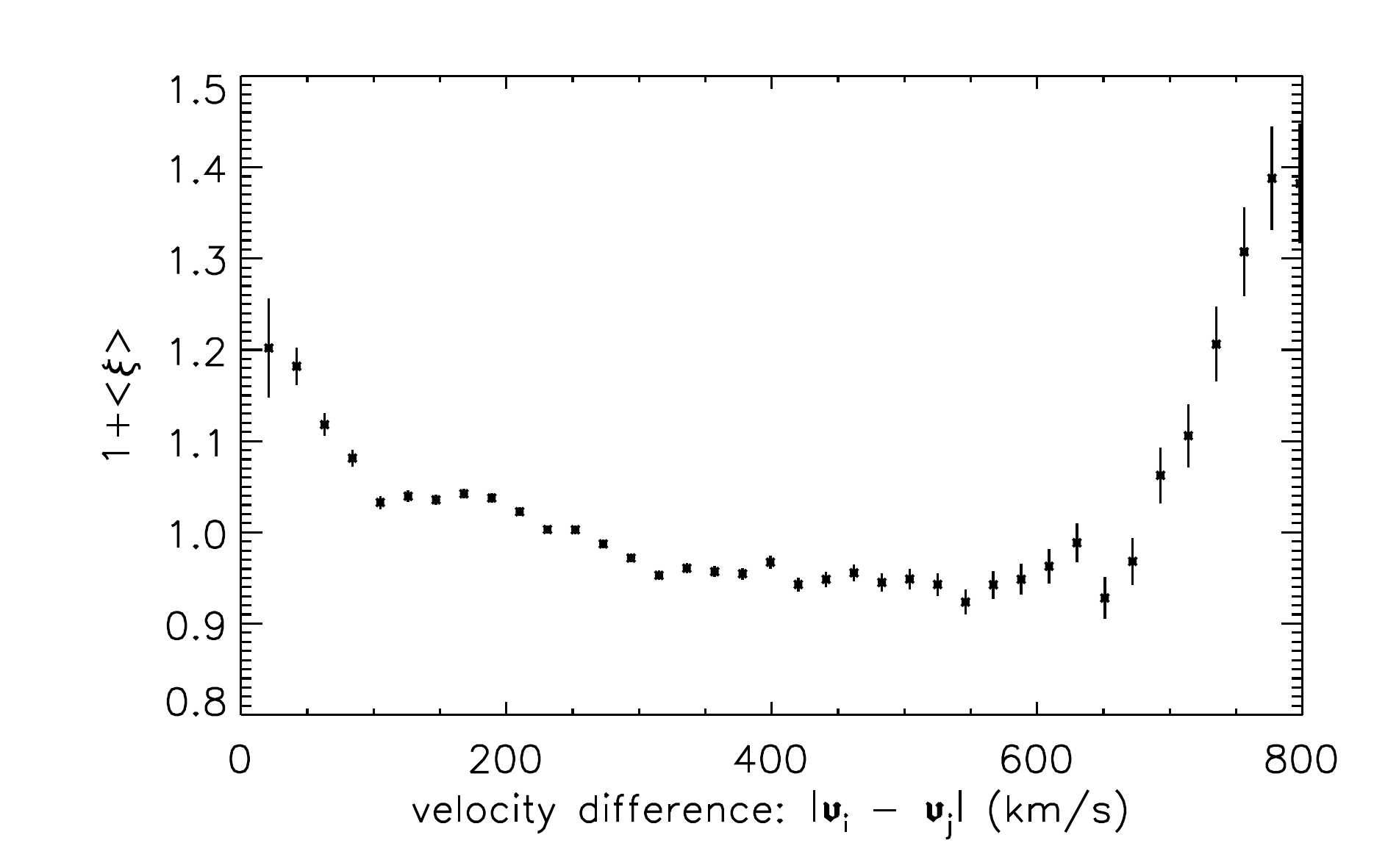} 
\caption{Amplitude of the velocity correlation function as a function of
  velocity separation defined as in Eq.~(\ref{eq:xi}). An excess of
  kinematic structure in our dataset compared to random (reshuffled)
  realisations is clearly apparent for small and for very large
  velocity separations.}
\label{fig:xi-tgas}
\end{figure}

Fig.~\ref{fig:xi-tgas} shows that the velocity correlation function
reveals an excess of structure on small scales well-above the signal
found in the randomised sets. The first few bins indicate a statistically
very significant excess of pairs. The error bars in this plot reflect
the Poisson statistics uncertainties. For example, the data shows 486
pairs with velocity separations $< 20$~km/s while in the random sets,
the average is 404.4, implying that in the first bin only there are 82
pairs of stars in excess. The significance level is $\sim 3.7\sigma$
assuming the uncertainty is Poissonian. Similarly, there is a very
significant excess ($8.8 \sigma$) of pairs for velocity separations
$20 \le \Delta < 40$~ km/s, with 3264 data pairs and 2761.5 random pairs on average.

If we focus on the first velocity bin, we can identify the stars most
likely to be related to the excess of pairs. We have found that 112
stars appear twice in a close pair, 59 stars 3 times, 24 stars 4
times, 10 stars are 5 times in tight pairs, and one star appears in 7
tight pairs. These results mean that the signal we detect with the
velocity correlation function is not due to a single stream or
structure. Furthermore, we have checked that the pairs are not
  due to binaries by computing the average physical separation between stars in
  the same tight ($\Delta < 20$~km/s) kinematic pair. We found this distance to be on
  average 0.95~kpc, with the closest stars in a pair separated on
  the sky by $\sim 6$~deg at a distance of $\sim 2.75$~kpc, implying a
  physical separation of $\sim 0.27$~kpc.

In addition, also note that at large velocity differences the correlation function
seems to indicate a signal well-above that expected from the
randomised sets. We shall see that this is plausibly related to an
excess of stars on retrograde orbits, since such large velocity
differences only can involve objects with extreme kinematics.

In the computation of the correlation function, we have not explicitly
  ``corrected'' for the effect of velocity errors. In general we expect that these
  will tend to lower the significance obtained, especially in the first velocity
  bin, i.e. the number of real pairs found with small velocity separations is likely a lower limit.  
In fact, the small change in slope in the correlation function seen in the first bin of 
Fig.~\ref{fig:xi-tgas} could potentially be related to  this effect.

\subsection{Distribution in Integrals of Motion space}
\label{sec:iom}

For each of the stars in our halo sample, we compute their angular momentum and their energy for a
potential consisting of a logarithmic halo, a Miyamoto-Nagai disk and
a Hernquist bulge: $\Phi = \Phi_{\rm halo} + \Phi_{\rm disk} +
\Phi_{\rm bulge}$, where
\begin{equation}
\label{sag_eq:halo}
\Phi_{\rm halo} = v^2_{\rm halo} \ln (1 + R^2/d^2 + z^2/d^2),
\end{equation}
with $v_{\rm halo} = 173.2 $~km/s, and $d = 12$ kpc,
\begin{equation}
\Phi_{\rm disk} = - \frac{G M_{\rm disk}}{\sqrt{R^2 + \left(a_{\rm d} +
\sqrt{z^2 + b_{\rm d}^2}\right)^2}},
\label{sag_eq:disk}
\end{equation}
with $M_{\rm disk} = 6.3 \times 10^{10} \sm $, $a_{\rm d}$ = 6.5 kpc
and $b_{\rm d}=0.26$ kpc, and
\begin{equation}
\Phi_{\rm bulge} = - \frac{G M_{\rm bulge}}{r + c_{\rm b}},
\label{sag_eq:bulge}
\end{equation}
with $M_{\rm bulge} = 2.1 \times 10^{10} \sm$ and $c_{\rm
  b}=0.7$~kpc. The numerical values of the relevant parameters in
these models are chosen to provide a reasonable fit to the rotation
curve of the Milky Way.  In practice the exact form of the potential
is not very relevant provided it represents a fair description in the
volume probed by the sample, as it acts mostly as a zero
point offset from the kinetic energy.

\begin{figure}
\centering
\includegraphics[width=8.5cm]{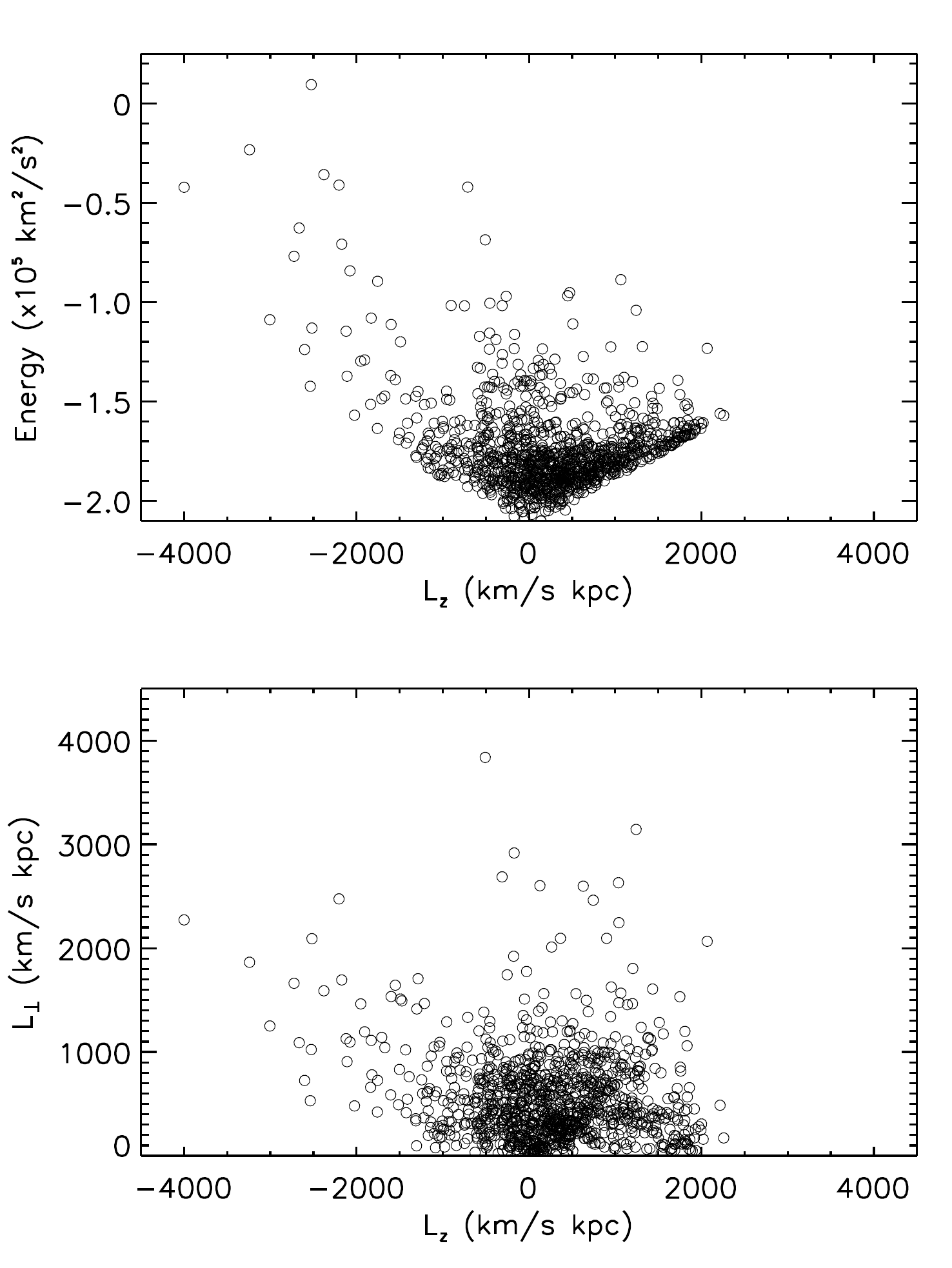} 
\caption{Distribution of Energy vs $L_z$ (top panel), and $L_\perp$ vs
  $L_z$ (bottom panel) for the halo metallicity selected sample obtained from the
  cross-match of TGAS and RAVE.}
\label{fig:iom}
\end{figure}

Figure \ref{fig:iom} shows the distribution in energy and $z$-angular
momentum $L_z$ on the top panel, and the $L_\perp =
\sqrt{L_x^2+L_y^2}$ vs $L_z$ on the bottom panel for our sample of
halo stars (including one random realisation of the ``hole'' disk
stars). The disk stars occupy a very distinct region in these spaces
(note the over-density at $L_z \sim 1800$~km/s~kpc), with most of the
halo having a much more extended distribution both in energy and
$L_z$. The stars with high binding energies ($E \lesssim -1.6 \times
10^5$~km$^2$/s$^2$) have little net angular momentum, but in contrast
most of the stars with lower binding energies, $E > -1.3 \times
10^5$~km$^2$/s$^2$ appear to have rather large retrograde motions. This
striking difference was never seen so prominently in a local sample of
stars, although hints of this behaviour can be found in many published
works as we discuss below.

\subsection{The less-bound halo: A retrograde component}
\label{sec:retro}

\subsubsection{The reality of the retrograde component}
\label{sec:retro-conv}

Figure \ref{fig:iom} shows evidence that an important fraction of the halo
stars with low binding energies that visit the Sun's vicinity (and are
therefore part of our sample) are on retrograde orbits. For example,
this percentage is 57.6\% for $E > -1.6 \times 10^5$~km$^2$/s$^2$, for
$E > -1.3 \times 10^5$~km$^2$/s$^2$ it is 72.7\%, while for $E > -1.2 \times
10^5$~km$^2$/s$^2$ the percentage is 84.9\%. At these low binding energies (lower
than that of the Sun), there appear to be two main features or plumes,
namely stars that are only slightly retrograde and which have $L_z
\sim -500$ km/s kpc, and stars with very retrograde orbits, with $L_z
< -1000$ km/s kpc.

Since there has been an important debate in the literature about the
presence of net retrograde rotation in the Galactic outer halo
\citep[e.g.][]{Carollo2007,Schoenrich2011,Beers2012}, it seems
relevant to determine whether the high proportion of stars in our
sample with low binding energies and with retrograde motions could be
an artefact of large distance and proper motion errors.  It does not
appear too unlikely that errors or even a wrong value of the circular
velocity of the Local Standard of Rest could shift slightly the plume with $L_z \sim
-500$ km/s kpc into the prograde region. For the very retrograde stars, however, this seems to
be less plausible.

\begin{figure}
\centering
\includegraphics[width=8.5cm]{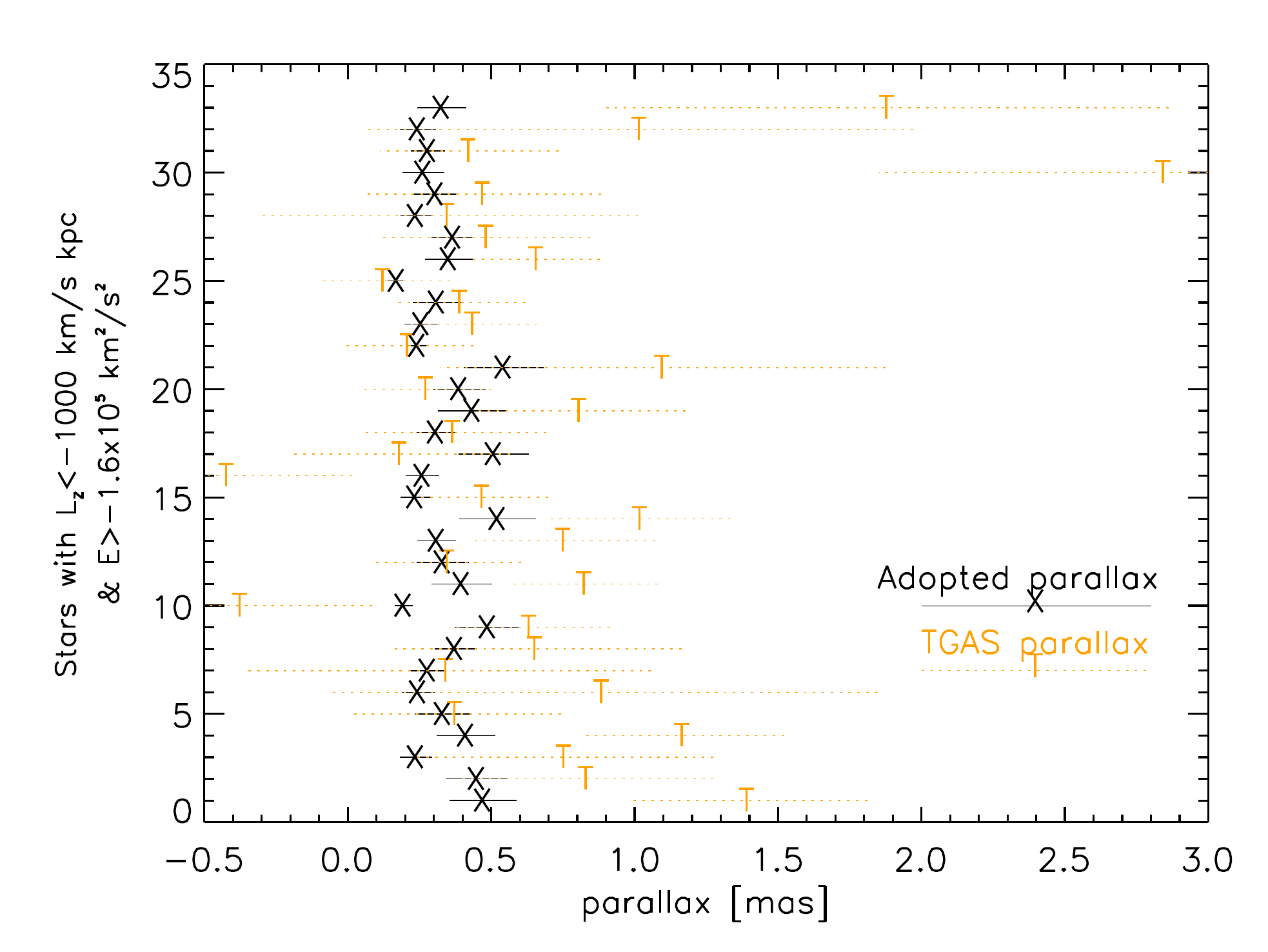} 
\caption{Comparison of the adopted parallaxes (defined as those which have
  the smallest relative error when comparing the TGAS and RAVE
  estimates for the high quality dataset defined as in
  Sec.~\ref{sec:TGAS}), and the TGAS parallaxes for the 33 stars with
  low binding energies and extremely retrograde motions.}
\label{fig:parallax_errors}
\end{figure}

Nonetheless, in Figure \ref{fig:parallax_errors} we compared the
parallaxes for TGAS and RAVE (scaled as described in
Sec.~\ref{sec:TGAS}) for the stars with $L_z < -1000$~km/s kpc and $E
> -1.6 \times 10^5$ km$^2$/s$^2$. This figure reassuringly shows that
the parallaxes derived from both datasets are consistent within the
errors for the vast majority of the stars. We focus on the parallaxes
because, even though large velocity errors can also be due to large
proper motion uncertainties, in our case they are driven mostly by the
parallax error, as effectively $\epsilon(v_y) \propto 4.74
\epsilon(\varpi) |\mu|$ because of the large magnitude of the proper
motion.

To quantify more directly the effect of uncertainties on the reality
of the retrograde halo component, we have convolved all observables
with their errors, and re-computed the distribution of stars in energy and
$L_z$ space. We find that in all the 1000 realisations made of the
data, and for different cuts in energy (with $E > -1.6 \times 10^5$
km$^2$/s$^2$) there is an excess of stars with negative $L_z$ of similar
amplitude as in the data. For example, for $E > -1.6 \times 10^5$
km$^2$/s$^2$, the fraction of stars in this region of ``Integrals of
Motion'' space, is 57.1\% $\pm$ 2.4\% (compared to 57.6\% in the data),
while for $E > -1.3 \times 10^5$ km$^2$/s$^2$ it is 71.8\% $\pm$ 4.2\%
(compared to 72.7\% in the data).  Therefore we may conclude the
errors alone cannot make the less bound halo become more prograde. In
other words, the signal we detect is not an artefact of the errors.

\begin{figure*}
\centering
\includegraphics[width=16cm]{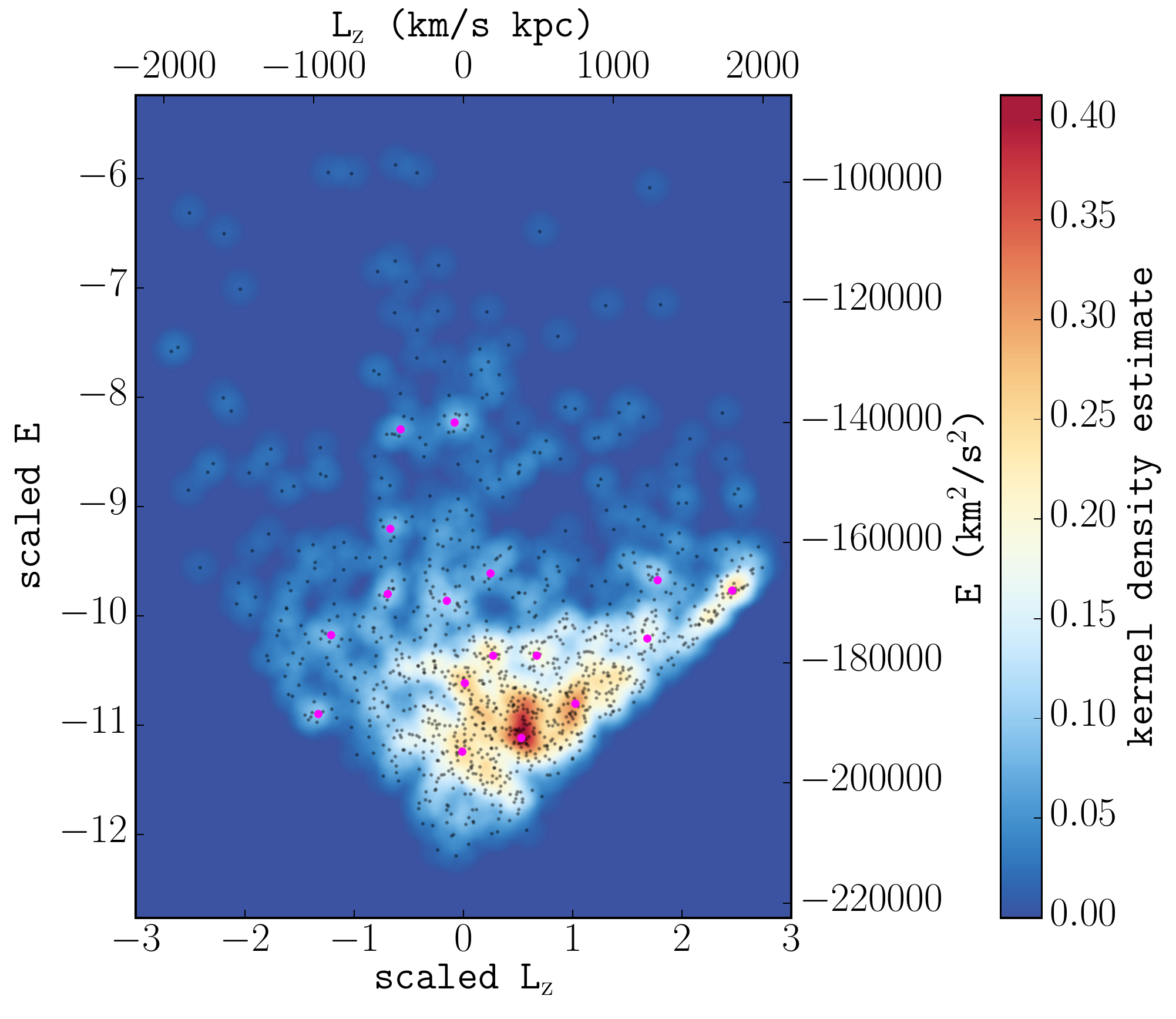}
\caption{Kernel density estimate of the distribution of our sample of
  halo stars in $E-L_z$ space. The stars themselves are shown as black
  dots.  The relative peaks of the density distribution are marked
  with solid magenta circles.}
\label{fig:kde}
\end{figure*}

\subsubsection{Significance}
\label{sec:retro_significance}

We may use the 1000 randomised (re-shuffled) realisations of the data
introduced in Sec.~\ref{sec:corr_f} to establish the statistical
significance of the retrograde component of the less-bound halo.  For
each realisation, we recompute the energies and angular momenta of the
stars using their reshuffled velocities. We then count the fraction of
bound stars $n_R$ above a given energy $E_{min}$ and with $L_z <
0$~km/s~kpc. We find that no realisation has as many stars as the data
in this region of ``Integrals of Motion'' space for values of $-1.6 \le
E_{min} \le -1.2 \times 10^5$ km$^2$/s$^2$. This implies that the
probability of finding the observed fraction of retrograde moving
stars is smaller than 1 in 1000, or 0.1\%.

The average $\langle n_R \rangle$ and standard deviation
$\sigma_{n_{R}}$ of the randomised datasets (which use the
same spatial distribution, and 1D velocity distributions as the data)
allows us to define a significance parameter, $s = (n_{obs} - \langle
n_R \rangle)/\sigma_{n{_R}}$. Depending on $E_{min}$, the minimum
energy considered, typical values of $s$ range from 4.2 to 7.7, again
indicating that the excess of loosely bound stars with retrograde
orbits in our sample is statistically very significant.

\subsection{The more bound halo: full of structure}
\label{sec:substr-iom}

We perform a comprehensive analysis of our sample of halo stars in the
space of ``Integrals of Motion'', now looking to identify
and characterise overdensities that could correspond to accreted satellites that have
contributed to the build-up of the stellar halo.

\subsubsection{Statistical analysis}

We first focus on $E-L_z$ space.  We constrain our analysis to the
most highly bound and populated region in the top panel of  Fig.~\ref{fig:iom},
selecting the stars that have $-2000 \le L_z \le 2000$~km/s~kpc, and
$-2.1\times10^5 \le E \le -1.0\times 10^5$~km$^2$/s$^2$. To determine
the density field of the stars in $E-L_z$ space, we apply the
\textsc{sci-kit learn} implementation of a non-parametric density
estimator that uses an Epanechnikov kernel. For optimal performance of
the kernel density estimator, we have scaled the data to unit
variance. We used the cross-validation method
\citep[e.g.][]{cross-validation}, also implemented in \textsc{sci-kit
  learn}, to determine the optimal bandwidth for the kernel density
estimator, and found it to be 0.2 in scaled units. The result of this
processing is shown in Fig.~\ref{fig:kde}. Since any of the easily
visible overdensities in Fig.~\ref{fig:kde} could be due to stars
that were once part of a single accreted object, we are interested in
determining their precise location.  To this end, we applied a maximum
filter in order to identify the relative peaks in the underlying
density distribution. We found 17 such local maxima, which are marked
in Fig.~\ref{fig:kde} with solid magenta circles.

\begin{figure*}
\centering
\includegraphics[height=8cm,clip=true]{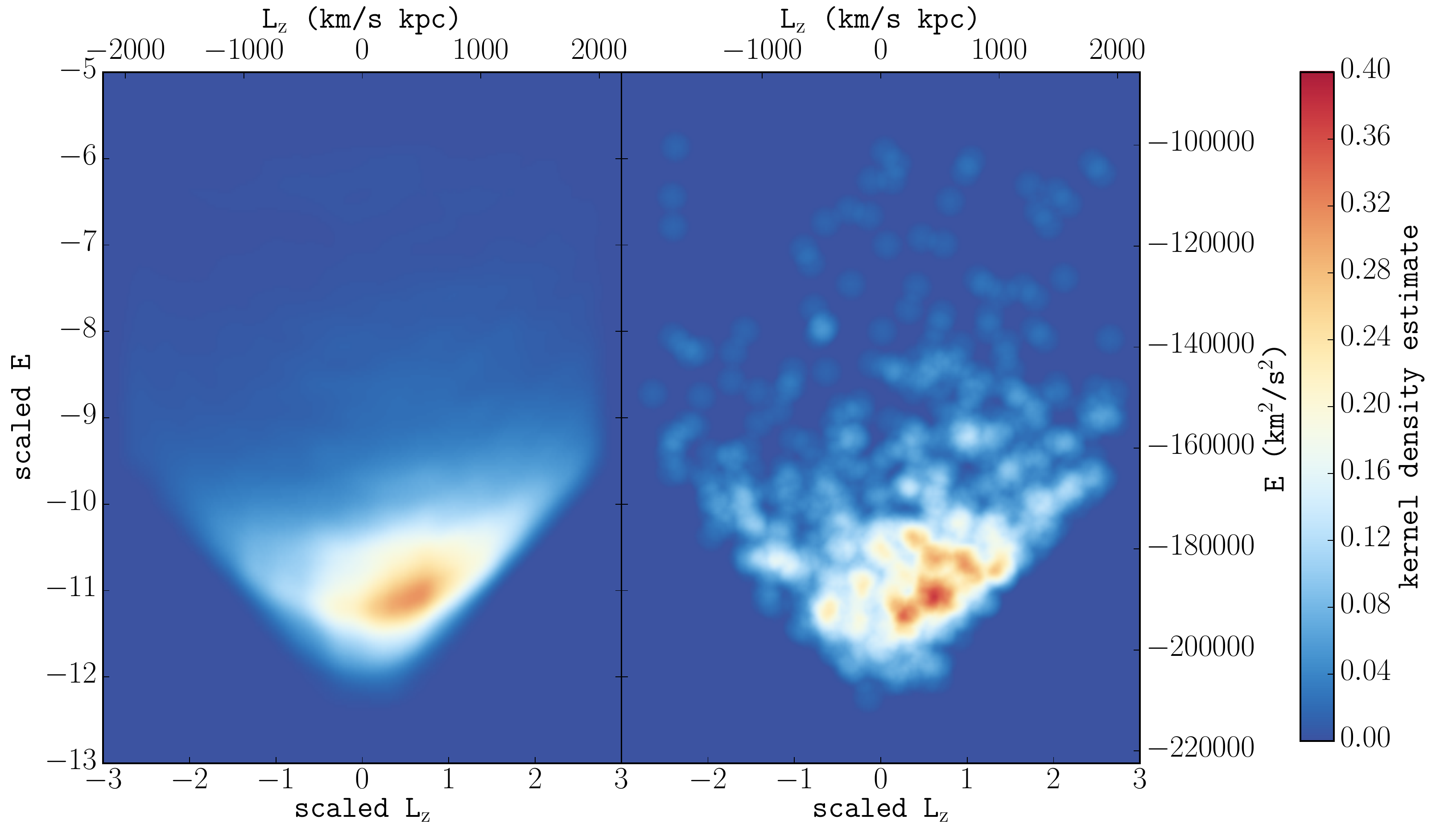} 
\caption{The left panel shows the average density of all 1000
  randomised realisations of the data in ``Integrals of Motion space'',
  using the same density estimator as for the real data. As an
  example, the right panel shows one of the random realisations.}
\label{fig:randomised}
\end{figure*}

\begin{figure*}
\centering
\includegraphics[height=8cm,clip=true]{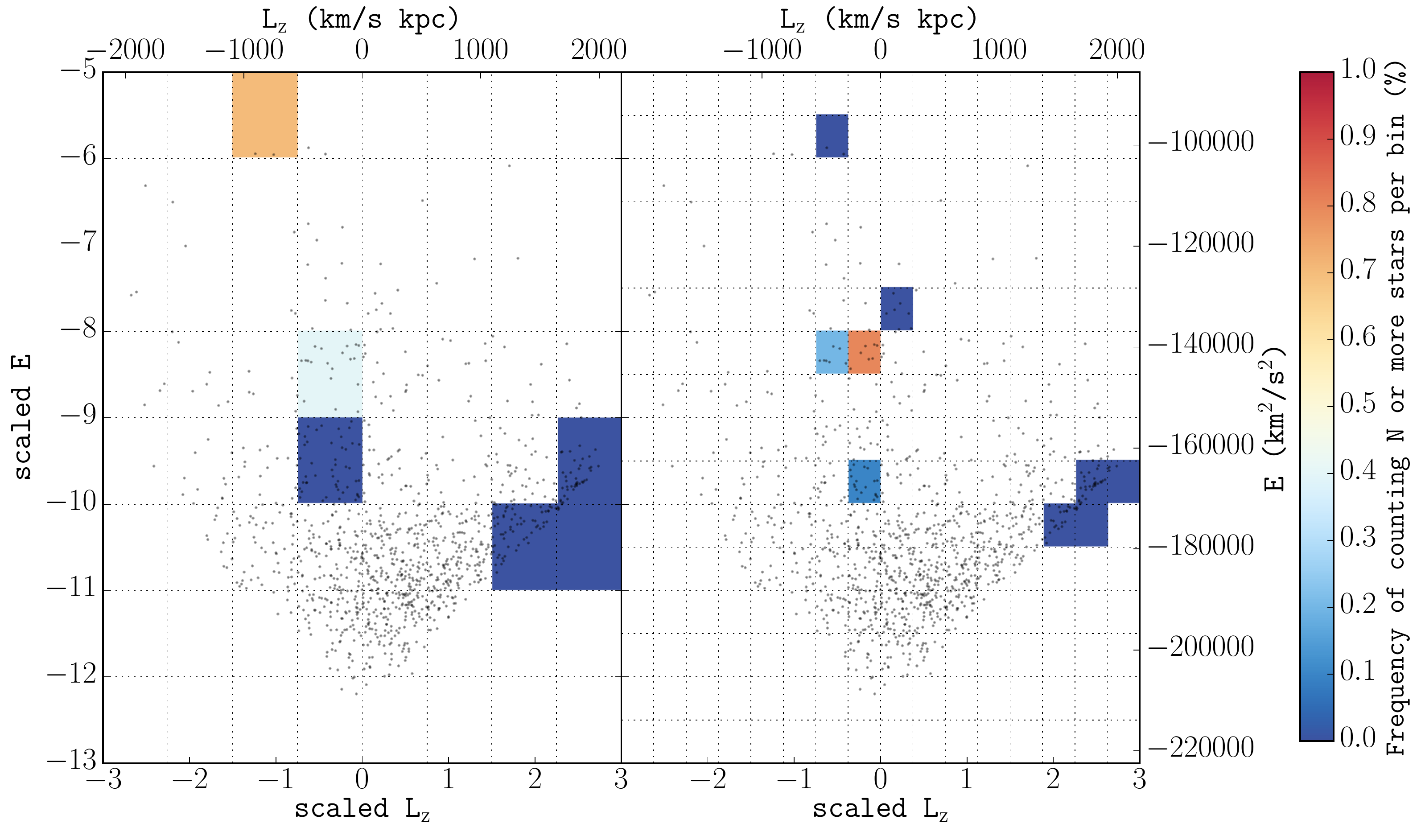} 
\caption{To determine the statistical significance of the
  overdensities we have identified in Fig.~\ref{fig:kde}, we bin the
  data in a 2D grid, and count how often we observe as many or more
  points in the randomised realisations compared to the real data
  set. For clarity, only the bins having a frequency of such
  occurrence of 1\% or less are coloured. The left panel shows a
  $8\times8$ grid, while the right one shows a $16\times16$ grid.}
\label{fig:gridstat}
\end{figure*}

\begin{figure*}
\centering
\includegraphics[width=16cm]{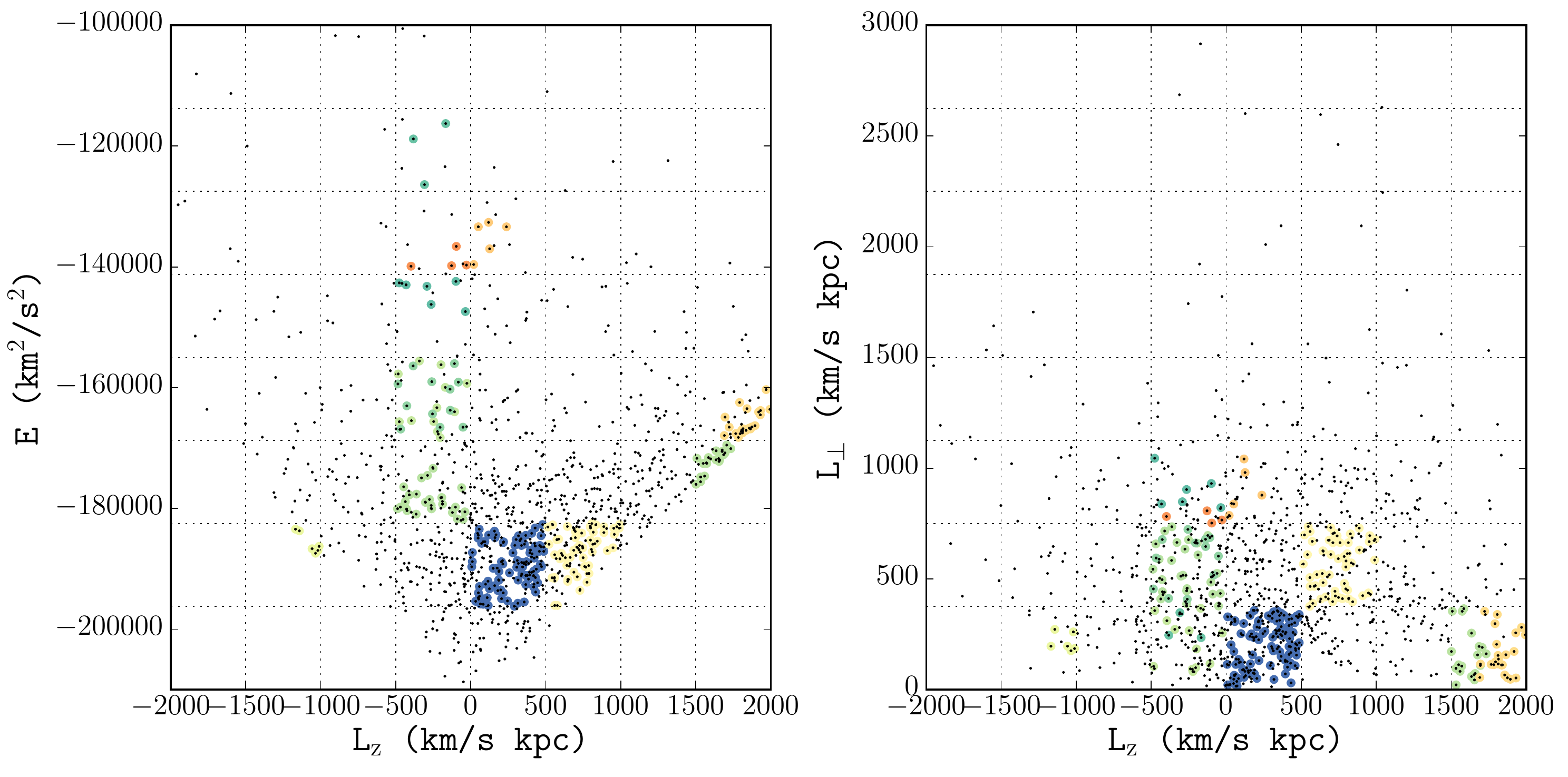} 
\caption{As in Fig.~\ref{fig:gridstat}, but now for the 3D full ``Integrals of
  Motion'' space, we determine the statistical significance of
  overdensities by counting how often we observe as many or more
  points in the randomised realisations compared to the real data
  set. Only stars falling in the bins having a frequency of such
  occurrence of 1\% or less are coloured.}
\label{fig:3D-poisson}
\end{figure*}

Examining the randomised datasets, an example of which is shown in the
right panel of Fig.~\ref{fig:randomised}, we see that their density
distribution is generally also not smooth, and that several clearly
distinct overdensities can be often discerned, especially
in the regions of higher binding energy. Therefore, we need to
investigate the probability that any of the overdensities we
determined in the real data may happen by chance. To do this, we bin
the data in $E-L_z$ space on a series of regular grids with
different bin sizes ($8\times8$, $16\times16$, $16\times16$,
$64\times64$). The different bin sizes are important as we want to
explore how the significance of the structures varies with scale.  For
each bin, we count how often we observe as many or more stars in the
randomised datasets compared to the real data. We then mark those
bins for which this frequency is 1\% or less. Examples of this
procedure are shown in Fig.~\ref{fig:gridstat}. We find 4 local maxima
identified in Fig.~\ref{fig:kde} to fall into bins that satisfy this
criteria, for all the grids explored.  When we perform a similar
analysis and compare a random realisation to the remainder 999 random
sets, we find that none depicts probability levels as low or lower
than 1\%, indicating that our strict significance cut removes false
positives. The fact that the overdensities
identified in the data are extremely hard to occur by chance and that they
appear independently of the grid used, makes us
confident that they are indeed due to genuine substructures in the
stellar halo of our Galaxy.

On the other hand, comparison of Fig.~\ref{fig:kde} and
Fig.~\ref{fig:gridstat} reveals that several of the peaks in $E-L_z$
space, particularly those located in the denser regions, do not appear
to be statistically very significant according to the above
analysis. To determine whether this could be due to a projection
effect, we perform an equivalent statistical analysis but now in 3D,
i.e. in $E-L_z-L_\perp$ space. As before, we divide the space in bins
of equal number in all directions. We then count the number of stars
in the real data and in the randomised datasets, and identify those
bins for which the frequency of finding as many or
more stars in the random sets as in the real data, is less than 1\%. The results of this
exercise are shown in Fig.~\ref{fig:3D-poisson} for the two
projections of ``Integrals of Motion'' space for the $8\times8\times8$
grid.
\begin{figure*}
\centering
\includegraphics[width=16cm]{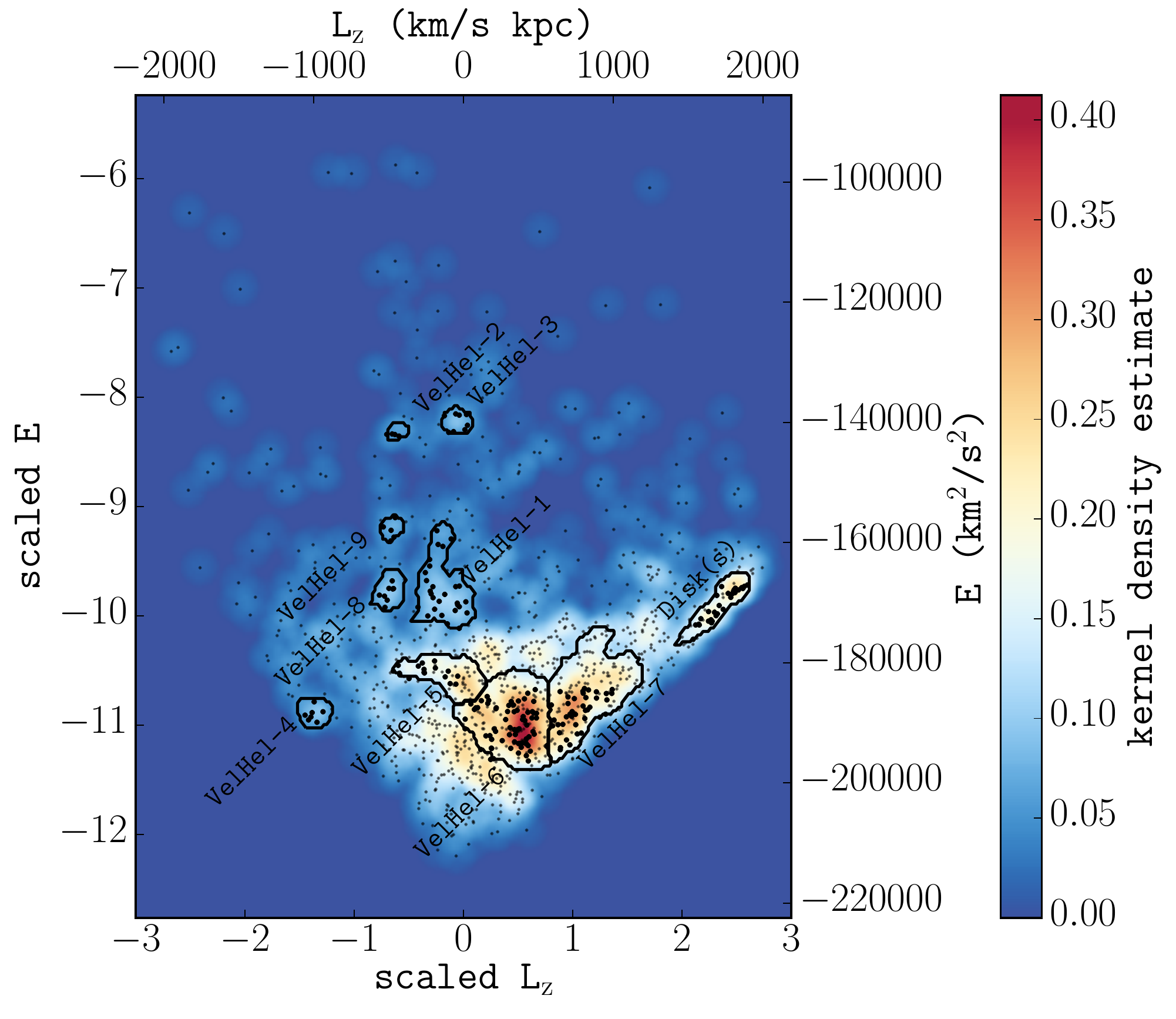} 
\caption{The black contours mark the extent of the substructures we
  have identified with the highest confidence to be real. These
  contours have been determined using a watershed algorithm but the
  member stars (indicated in this figure by the solid dots) is
  determined by considering also the results of the 3D Poisson
  analysis.  For the structures in the less crowded regions of the
  $E-L_z$ space, we have set the watershed level to 0.05 of the kernel
  density estimate, while for the densest parts we set the level to 0.13.  We
  find that with these values we best trace the significant structures outlined
  by the density map.}
\label{fig:watershed}
\end{figure*}
\begin{figure*}
\centering
\includegraphics[width=8.5cm]{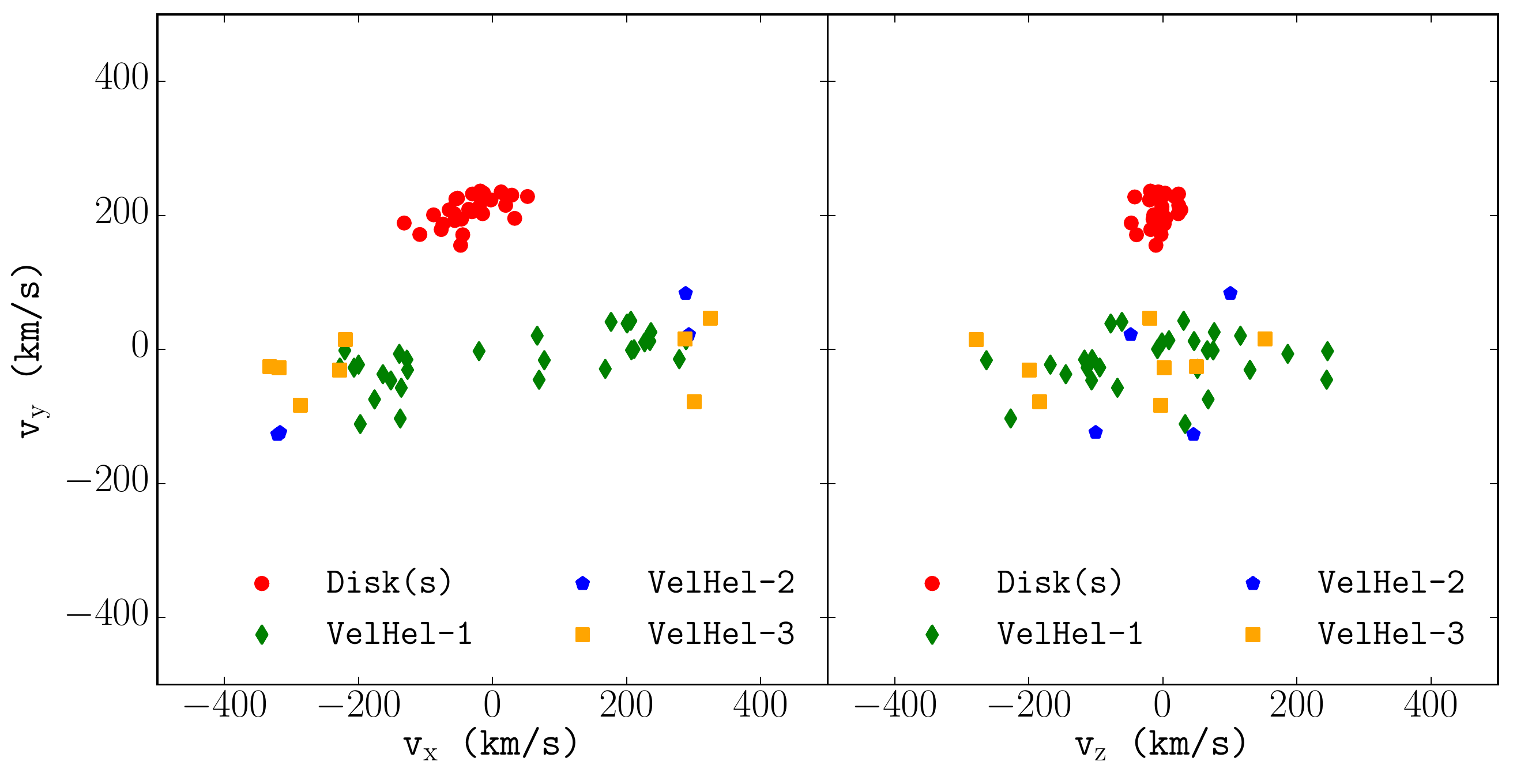} 
\includegraphics[width=8.5cm]{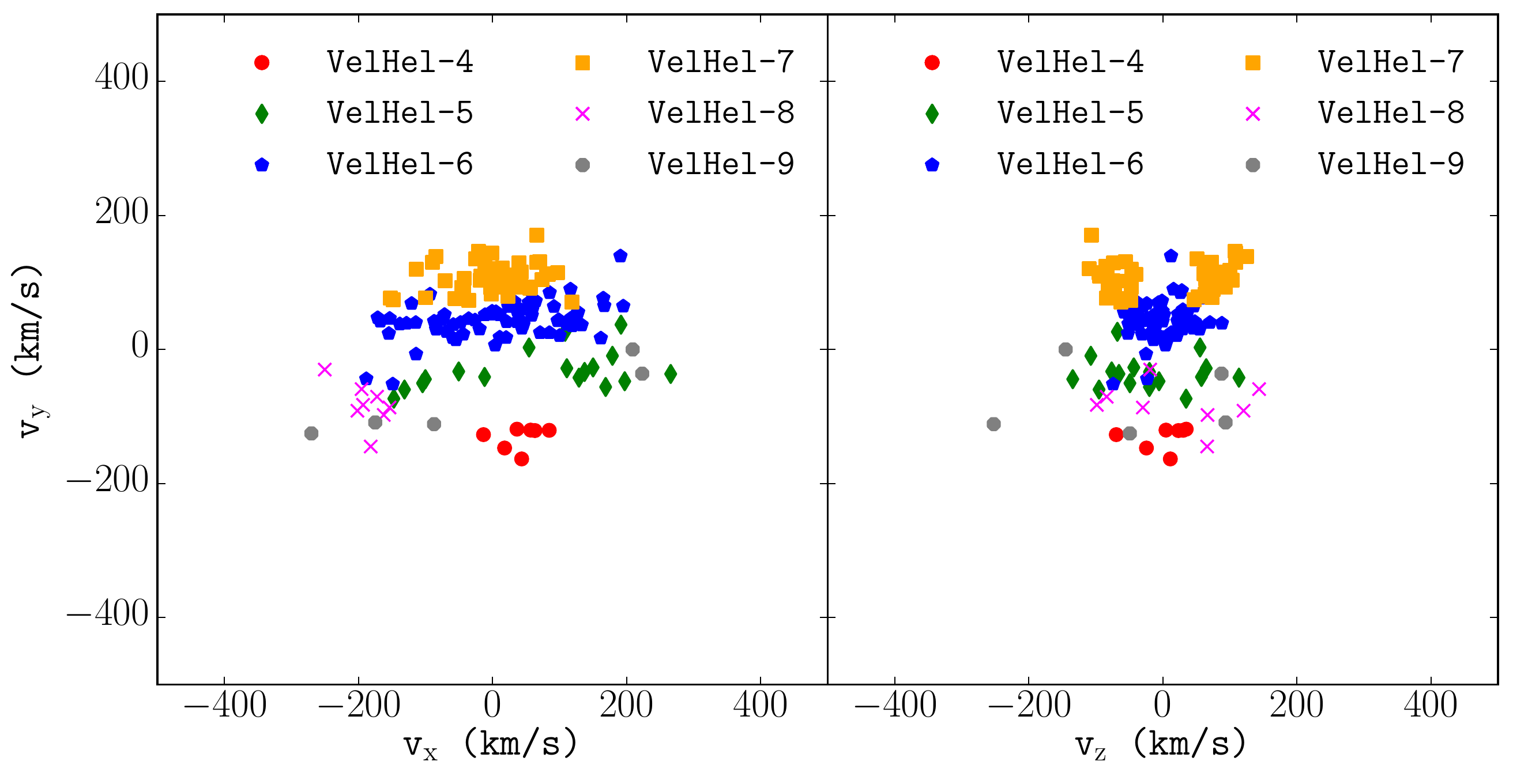}
\caption{Cartesian velocities for the stars comprising the identified structures.
The velocities are not corrected for the effects of the Solar motion.}
\label{fig:vxvyvz}
\end{figure*}

Fig.~\ref{fig:3D-poisson} confirms the statistical significance of the
bins identified in our 2D analysis. It also helps to isolate better
the stars belonging to the structures by using the third dimension,
$L_\perp$. On the other hand, several more regions are now
identified. In fact, all these regions have a good correspondance with a subset of
the structures visible in the 2D density field of the $E-L_z$ space shown
in Fig.~\ref{fig:kde}. Of the 17 maxima we had discerned in this
density field, a total of 10 appear to be associated to true statistically
significant overdensities of stars in ``Integrals of Motion'' space.
\begin{figure*}
\centering
\includegraphics[width=14cm]{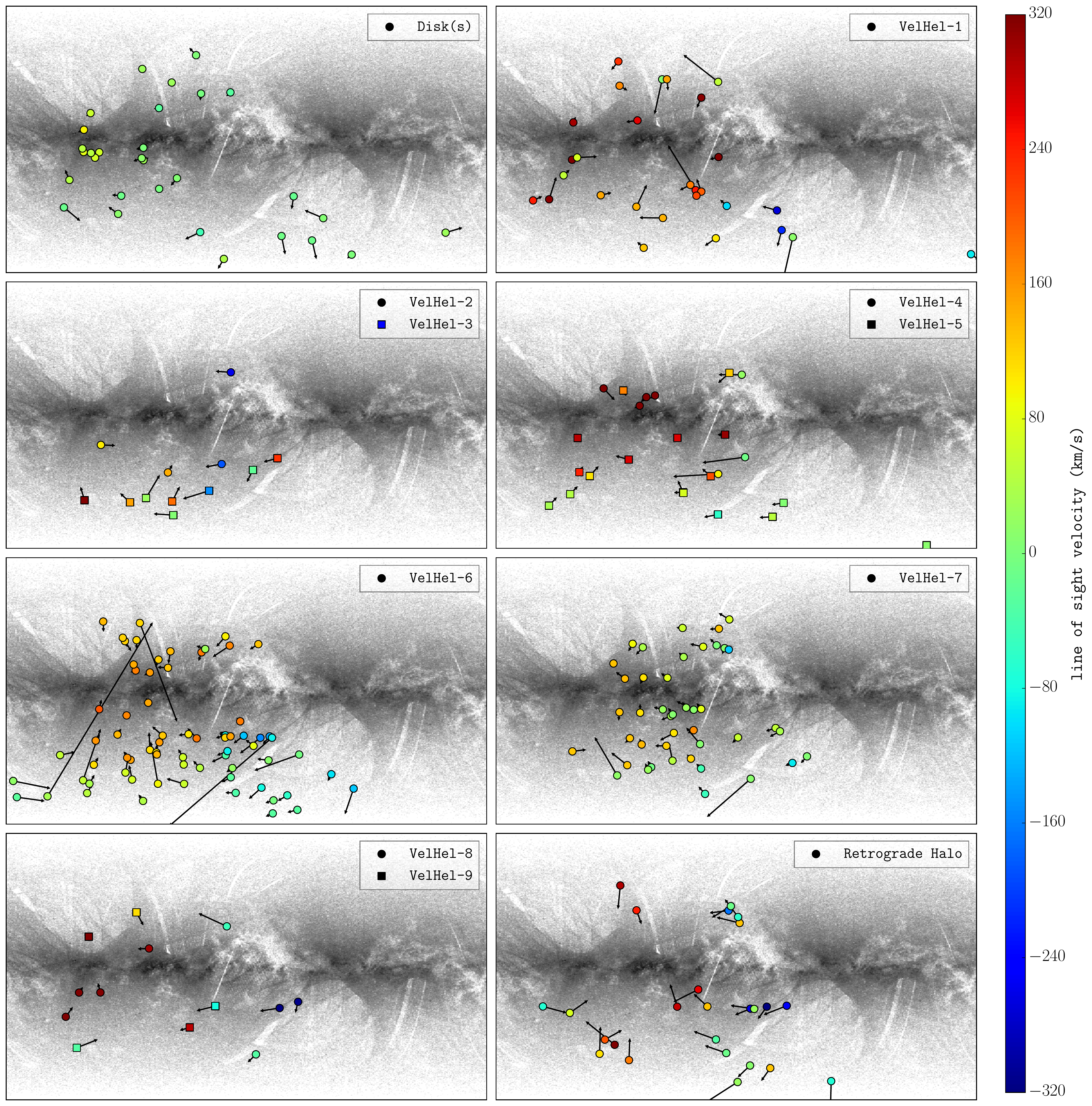} 
\caption{Sky distribution in Galactic coordinates $(l,b)$ of the stars in the
  structures identified in Sec.~\ref{sec:id}. The arrows indicate
  their velocities in the Galactic latitude and longitude,
  while the colour corresponds to their line-of-sight velocities. In
  the background, the stars from the full TGAS release are shown.}
\label{fig:skymap}
\end{figure*}

\subsubsection{Characterisation of the substructures}
\label{sec:id}

Having identified which of the substructures in the $E-L_z$ space were
significant, we use the {\sc watershed} algorithm \citep{watershed} to
estimate their extent and to determine their constituent stars, as
shown in Fig.~\ref{fig:watershed}. This algorithm works by ``inverting'' 
the terrain (in this case taking the negative of the density distribution, i.e. $-\rho_{KDE}$), and uses the local minima 
(maxima in the density distribution) as sources of water, ``flooding'' the 
basins (structures) until a particular ``water level'' is reached, effectively 
determining the extents of those basins (overdensities in our case).

The 3D Poissonian analysis is particularly helpful in the densest
regions of the $E-L_z$, as those are most affected by contamination,
and there $L_\perp$ is crucial to disentangle membership. We use this
information to supplement the watershed algorithm and remove
``interlopers''. We consider as interlopers those stars that do not fall
in a significant bin in the 3D analysis, but that in projection are
located inside a specific contour of the watershed.

In the Appendix~\ref{App:starlist} we have tabulated the
positions and Tycho IDs of the stars, members of each of the 10
substructures we have identified with the above analysis. The most
prominent, and statistically significant of these 10 substructures,
located at $E \sim -1.7 \times 10^5$~km$^2$/s$^2$ and $L_z \sim
1800$~km/s~kpc (-10 and 2.5 in scaled units, respectively), is in fact
due to the disk (contamination) in our halo sample. Most of the rest are
previously unknown structures, and we dub them VelHel-1 through 9.

Fig.~\ref{fig:vxvyvz} shows the Cartesian velocities, not corrected
for the Solar motion, for the stars comprising the 9 structures and
for the overdensity associated to the disk.  The disk stars are easily
recognisable, having $v_y \sim$ 220~km/s. The rest of the substructures
are clustered to different degrees in velocity space, where they
sometimes define a single clump as for e.g. VelHel-4, VelHel-8 in
$v_x-v_y$ space, or distinct separate clumps or streams, as for
e.g. VelHel-7 in $v_y-v_z$ space. For such cases, it is likely that we
are probing different portions of the orbits of now dispersed accreted
satellites. This means that the debris is wrapped multiple times around
the Galaxy in order to transverse the same volume with a different velocity.

Fig.~\ref{fig:skymap} shows a projection of proper motion vectors onto 
the sky for the stars in the different groups as well as for the
stars in the very retrograde component (defined in this case as those 21 objects 
that have $L_z \le -1000$~km/s~kpc and $E > -1.4\times 10^5$~km/s~kpc). In
this figure, the arrows indicate the velocity vector corresponding to
the Galactic latitude and longitude directions, with components $v_b$
and $v_l$, while the color denotes the amplitude of the stars'
line-of-sight velocities. The background image shows the sky
distribution of all TGAS stars. Note that the stars in the various
groups are distributed over wide regions on the sky, and that single
kinematic clumps are rarely apparent in this projection. However,
several flows can often be seen, each possibly indicating a different stream from the
various overdensities identified in ``Integrals of Motion'' space.

\section{Discussion}\label{sec:disc}

\subsection{Data caveats and the robustness of the results}

We have found evidence of significant overdensities possibly
associated with merger debris in a sample of halo stars that was
selected on the basis of the RAVE metallicity determination. In the
derivation of the phase-space coordinates of this sample, we used the
parallaxes from TGAS or RAVE, depending on which had the smallest
relative error.  We have explored the impact of using the
absolute error instead  instead of the relative error as a discriminator, 
or of not scaling the RAVE parallaxes by 11\% for the giant stars, and found our 
results to be robust. Our relative error tolerance of 30\% may be considered
relatively large, but this is necessary to have a sufficiently large
sample of stars in which to identify the subtle substructures the
models predict. Nonetheless, with stricter error cuts, the global kinematic 
properties remain similar, and the retrograde halo component, although populated 
by fewer stars, is still clearly visible.

To explore the effect of the uncertainties in energy and $L_z$
  on the substructures reported in
  Section~\ref{sec:substr-iom}, we have used the 1000 samples
  introduced in Sec.~\ref{sec:retro-conv} that resulted from (re)convolving the
  observables with their errors.  We have ran our kernel density
  estimator on the resulting distributions of energy and $L_z$ with
  identical settings as for the real data.  As one could expect, we
  find that errors tend to slightly blur the structures, but generally
  not enough to make significant changes. Exceptions are those
  substructures containing few stars, where the effect of Poisson
  statistics is significant.

Since TGAS does not contain all TYCHO-2 stars, we also investigated
the phase-space distribution of all the TYCHO-2 stars included in the RAVE
sample, and found no important differences. Again, the retrograde
component is clearly apparent, and some of the other overdensities are
visible as well in this more complete, but less accurate dataset.

We have found $\sim 240$ stars\footnote{This includes 219 stars in the
10 structures identified in Sec.~\ref{sec:substr-iom} and 21 stars that are 
part of the very retrograde less-bound
component, defined here as $L_z < -1000$ km/s kpc and $E_{min} > -1.4 \times
10^5$~km$^2$/s$^2$.} in a sample of 1116 halo stars to be part of
overdensities in ``Integrals of Motion'' space. This is close to the
total number of stars found in tight velocity pairs (with velocity
separations smaller than 20 km/s) according to the velocity
correlation function analysis. In fact, there is a relatively good
correspondance between the stars located in the overdensities in the
densest part of the $E-L_z$ space and those in the tight kinematic
pairs.

It could be tempting to conclude, based on these numbers, that at
least $\sim 21\%$ of the stars in the halo near the Sun have therefore
been accreted. However, this conclusion would have to rely on a good
understanding of the completeness of our sample. This is in fact
difficult to determine at this point, both because of the selection
function of TGAS, especially at the bright end, and that of RAVE. However,
\citet{Wojno2016} and \citet[][]{Kunder2016} show that there
are no indications of any biases in the velocity
distributions of RAVE stars. Nonetheless, the determination of the true (relative)
contributions of the various overdensities to the stellar halo near
the Sun remains difficult to estimate reliably, even if only because the RAVE
survey has solely obtained data for stars visible from the southern
hemisphere. We will have to defer such analyses to future work; in
particular the second Gaia data release will make this possible.

\begin{figure}
\centering
\includegraphics[width=8.5cm,clip=true]{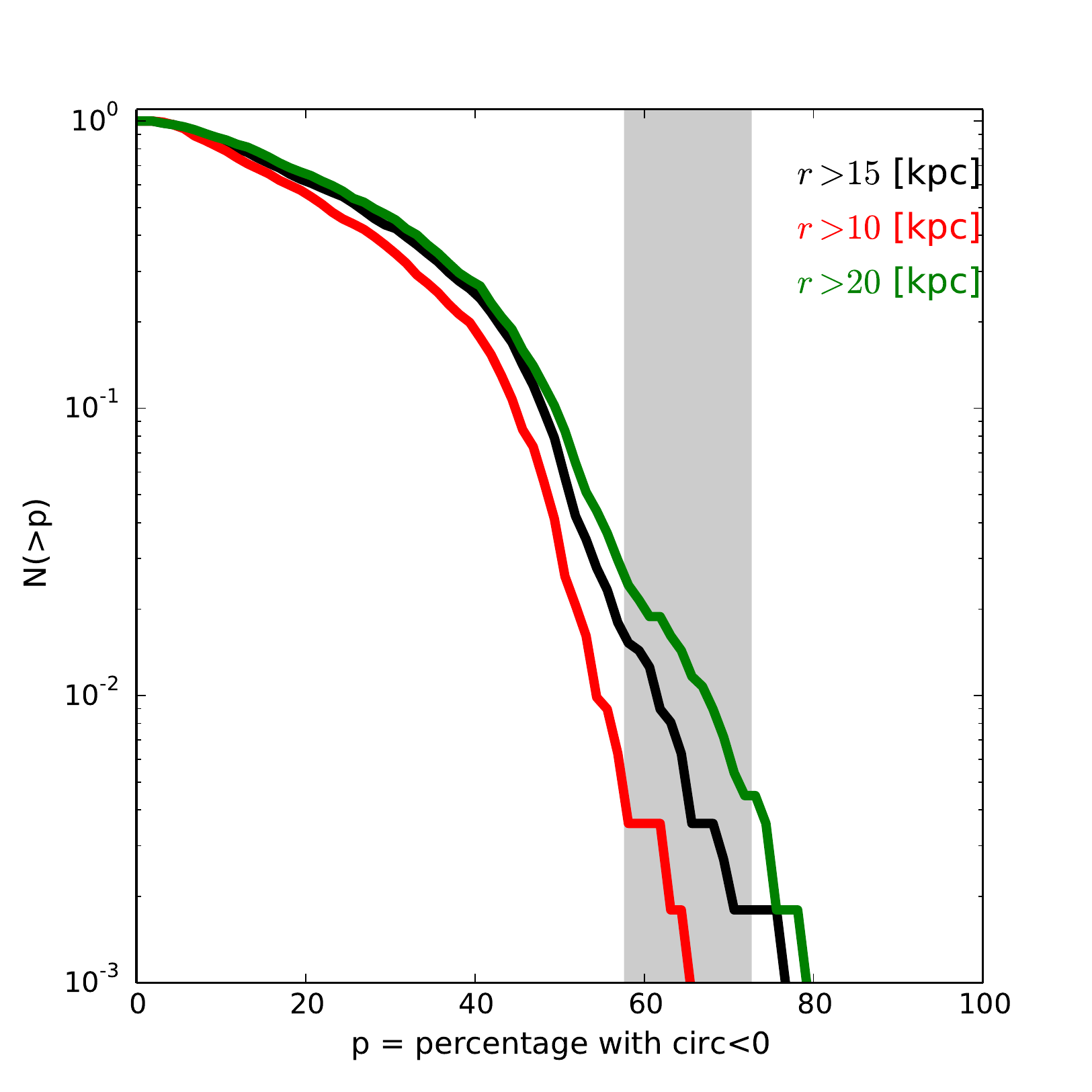} 
\caption{Cumulative fraction of Milky Way-mass galaxies in the Illustris simulation, containing a given
  fraction of retrograde halo stars beyond a radial distance of 10, 15
  and 20 kpc as indicated in the inset. The grey region denotes our
  estimated fraction of less-bound halo stars with retrograde motions
  in the real data.}
\label{fig:illustris}
\end{figure}

\subsection{The retrograde halo: $\omega$ Cen and more}

A literature search quickly demonstrates that there are independent
reports from many different surveys of predominantly retrograde
motions in (some portions of) the Galactic halo, indicating that this
is an important (sub)component
\citep{Carney1996,Carollo2007,NS2010,Majewski2012,An2015}. In our
sample, we have defined as less bound stars those with energies
$E_{min} > -1.6 \times 10^5$~km$^2$/s$^2$, for which we find 95
candidates, while if $E_{min} > -1.3 \times 10^5$~km$^2$/s$^2$, there
are 32 objects. What is remarkable, is that {\it nearly all} nearby halo stars
somewhat less bound than the Sun are on
retrograde orbits.

As Fig.~\ref{fig:iom} shows, this portion of the halo seems to contain
two features, stars that are only slightly retrograde, and some that
have very negative angular momenta. Those that are only slightly
retrograde, including the structure we have identified as VelHel-1
(and possibly also VelHel-3 and VelHel-5, see also
Fig.~\ref{fig:3D-poisson}) could perhaps be all associated to the
progenitor galaxy of $\omega$Cen \citep[see also][]{FT2015}.
\citet{Dinescu2002} built a good case for a scenario in which
$\omega$Cen represents the core of an accreted nucleated dwarf,
although she focused on debris on slightly tighter orbits, closer to
those defined by $\omega$Cen itself 
\citep[see also][]{Brook2003}. \citet{BF} and \citet{Tsuchiya}
used numerical simulations to convincingly argue that the progenitor
galaxy must have been a lot more massive if it had to sink in via
dynamical friction on a retrograde orbit. Such models predict a
trail of debris with a range of binding energies possibly resembling
the elongated feature we have identified in our analyses.

Although the prominence of $\omega$Cen has been put forward before, we have found that
it is not the only important contributor to the retrograde portion of the
halo. In this paper we have shown the striking dominance of halo stars on
less-bound retrograde orbits crossing the Sun's vicinity. Depending on
the minimum energy considered, between $\sim 58\%$ and 73\% of these
stars have motions that are significantly retrograde. Such a high
fraction appears to be intuitively somewhat surprising, and we
turn to cosmological simulations to establish how likely this is.

We use the Illustris cosmological hydrodynamical simulation
\citep{Vogelsberger2014a,Vogelsberger2014b,Nelson2015} to quantify the likelihood
of finding galaxies with a given fraction of their halo stars on
counterrotating orbits. To this end, we select all central galaxies
within the virial mass range $M_{200}=0.8 - 2.0 \times 10^{12}\; \rm
M_\odot$, which encompasses current estimates for mass of the Milky Way. We rotate each system
such that the total angular momentum of the galaxy
(defined by all the stars within $r_{\rm gal}=0.15 r_{200}$) points along the $z$-axis. In
this rotated system, we compute the fraction of stars outside a given
radial distance ($r>10, \; 15, \; 20$ kpc) that have negative angular
momentum in the $z$-direction, i.e. that have retrograde orbits.  The
results are shown in Fig.~\ref{fig:illustris}, which plots the
cumulative fraction of galaxies in our sample where the counter
rotating stellar halo accounts for more than a given fraction of the
total stellar mass outside the given radius. We find that $\sim$1\%
of the galaxies with Milky Way-like mass have a halo with more than $60\%
$ of the stars being retrograde outside $r=15$ kpc, and that about
half the sample has a counterrotating fraction equal or larger than
$\sim 30\%$.  For comparison, the values inferred from our halo sample
are shown with the gray shaded region. Such a large fraction of
retrograde halo stars thus seems rather unusual according to these
cosmological simulations, but is certainly not
impossible. In addition, it should be born in mind that our
sample is not complete, and that the fractions we have found could
have been overestimated because of selection effects. Furthermore,
because of numerical resolution limitations in the simulations, we
based our comparison on stellar particles located beyond a given
radius, while our observational selection is based on an energy
threshold for stars crossing the Solar neighbourhood now. Therefore
these conclusions should be taken as intriguing and suggestive, and
will need to be confirmed once the selection function of our sample is
better known, and should be contrasted to higher resolution cosmological
simulations.

\subsection{Comparison to cosmological simulations: granularity}

Another interesting point is to establish whether the level of
velocity granularity that we find in our dataset is consistent with
cosmological simulations of the formation of stellar halos. Because we
are not concerned with a specific sense of rotation with respect to a
major Galactic component, we may use the stellar halos that result
from coupling a semi-analytic galaxy formation model to the Aquarius
dark matter only simulations \citep{cooper2010}, after applying a
suitable tagging and resampling scheme to the dark matter particles
\citep{Lowing2015}. This suite of simulations therefore only models
the accreted component of a halo but has much higher resolution than
e.g. the Illustris cosmological hydrodynamical simulations. The
resulting Aquarius stellar halos have a range of stellar masses from
3.8 to 18.5$\times 10^8 \sm$.  Therefore to make a more direct
comparison to our own Galactic halo, we have scaled the simulated
stellar halos to have a stellar mass of $2.64\times 10^8\sm$,
i.e. that estimated for the Milky Way by \citet{Robin2003}. We do this
by downsampling the number of stars per population (e.g. main
sequence, red giant branch, etc) by the corresponding factors.

For each of the halos, we place a small sphere of 2.8 kpc radius at 8
kpc from the center, and randomly select 1116 objects amongst the
``simulated stars'' with magnitudes in the range $8 \le G \le
12.5$. We have applied the appropriate photometric band
transformations \citep{Jordi2010}, as the \citet{Lowing2015} dataset
provides the stellar magnitudes in the SDSS filters. We then compute the
velocity correlation function as we have done for our observed dataset,
and compared them to randomly (reshuffled) distributions, again as for the real data.

\begin{figure}
\centering
\includegraphics[width=9cm,clip=true]{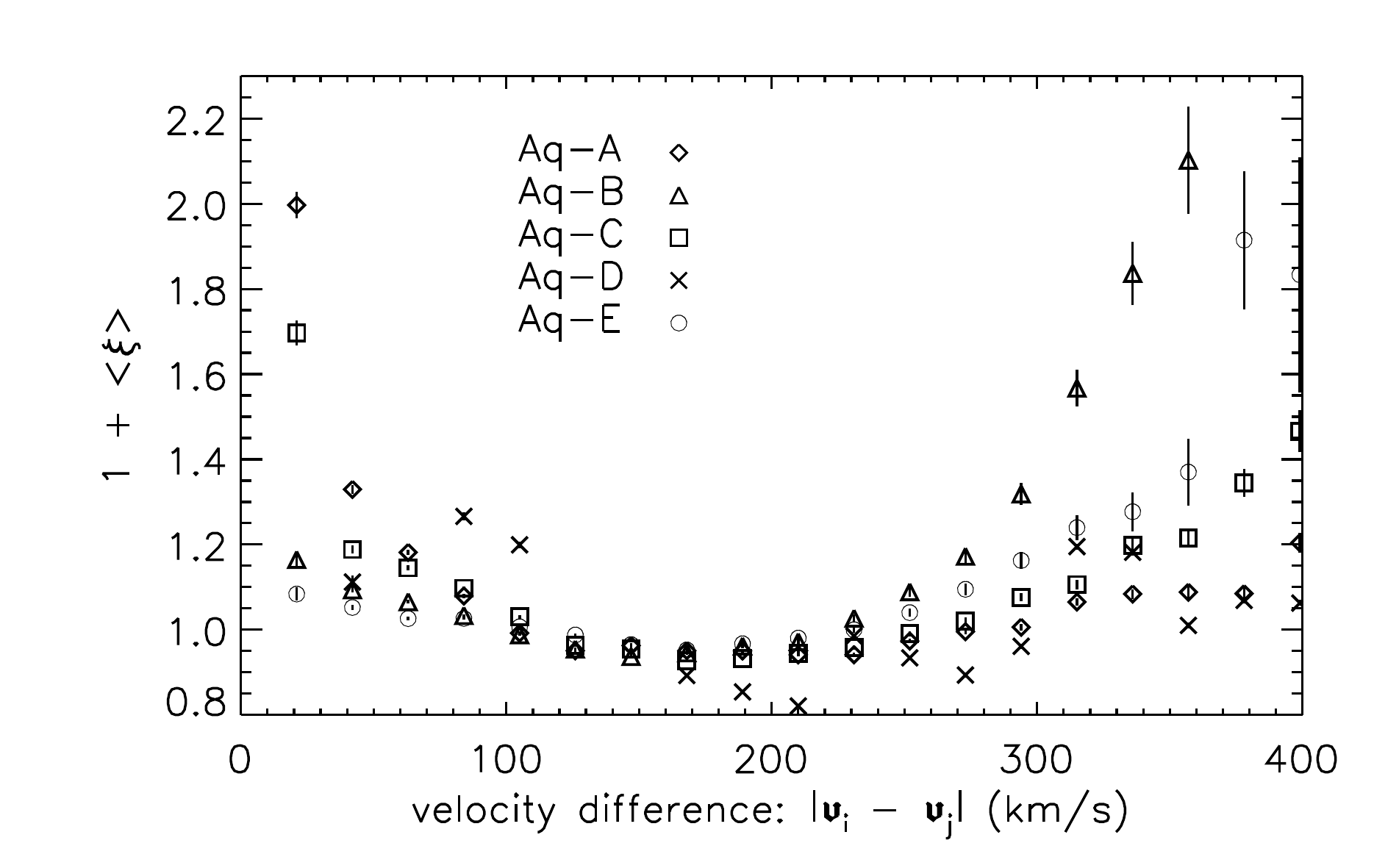} 
\caption{Velocity correlation function for a sample of ``stars''
  extracted from the stellar halos in the Aquarius simulations \citep{Lowing2015}. The
  stars are located in a sphere of 2.8 kpc and have magnitudes in the
  range $8 \le G \le 12.5$. The sample size is the same as our
  observed halo dataset. A comparison to Fig.~\ref{fig:xi-tgas} shows a
  similar excess of pairs in the simulations as in the data, although
  there is some variance in the simulations.}
\label{fig:aq-xi}
\end{figure}

In Figure \ref{fig:aq-xi} we show the resulting velocity correlation
functions. Interestingly, the amplitude of the signal in the simulated
halos is very comparable to the signal we find in our sample of halo
stars, although there is some scatter from simulation to
simulation. Note as well that the range of velocities over which we
make the comparison is different to that used for the data
and this is because the Aquarius halos have a smaller velocity
dispersion as they do not have a disk component. This means that the
maximum velocity separation is bound to be smaller. Nonetheless the
excess of pairs with large velocity separation is similar in amplitude and shape to
what we found in the data.

Taken at face value this comparison implies that the amount of
substructure present in our sample of Milky Way halo stars is
consistent to that expected for halos fully built via accretion. In
order to see significant differences, and to determine robustly the true relative
fractions of accreted vs in-situ halo stars would require a much
larger sample. Fortunately such a sample will become
available with the 2nd and subsequent data releases from the \Gaia
mission.

\subsection{New and old substructures}

We now review our knowledge of substructures in the space of ``Integrals
of Motion''. Firstly, it is important to note that the fact that the
region occupied by disk(s) appears as such a prominent overdensity in
our analyses would suggest that the disks have a rather extended
metal-poor tail \citep[see e.g.][]{Kordo2013}.  We have already
discussed the possible relation of structures VelHel-3, VelHel-1 and
VelHel-5 to $\omega$Cen debris. Perhaps VelHel-2, VelHel-8,
VelHel-9 are also associated to this, although they could also be
 independent structures themselves. On the other hand, structures
VelHel-6 and VelHel-7 were previously unknown, as well as
VelHel-4. This particular clump has very interesting kinematics as it
is on a low inclination orbit and rotates almost as fast as the disk but
in the opposite direction.

All these substructures are new and have not been reported in the
literature to the best of our knowledge. We have quickly inspected
their distribution in [$\alpha$/Fe] vs [Fe/H] as determined by the
RAVE survey pipeline and found them fairly undistinguishable from canonical
halo stars, and without any particular degree of clustering in this
chemical abundance space.

In comparison to other reports of substructure, such as e.g. based on
the SDSS survey by \citet[][]{Smith2009,ReF2015}, we do not find
overlap. Also noticeable is the absence of an overdensity of stars
associated to the \citet{h99} streams, although there is a small
kinematic group in our sample at the expected location, but which is
not picked up as statistically significant by our analysis. We have
inspected the distribution of these stars and found them to be
preferentially located at high Galactic latitude. The footprint of the
RAVE survey thus seems to hamper the discovery of more members, whose
presence has been confirmed for example in the SDSS survey, which does
cover well the north Galactic pole region. Similarly, the
overdensities reported by \citet{Smith2009,ReF2015} do not appear in
our sample, and these correspond to structures with rather high
$L_\perp$. Stars with large $L_\perp$ for a given $L_z$ have high
inclination orbits, and therefore move fast in the vertical direction
when crossing the Solar neighbourhood. They would be more easily
detected when observing towards the Galactic poles, but the RAVE
survey penalises these regions, although in the case of the south
Galactic pole, it is an issue of completeness with magnitude \citep{Wojno2016}. Selection effects such as these are clearly
important when estimating the fraction of the halo that is in
substructures, and require careful attention when making quantitative
assessments.

\section{Conclusions}
\label{sec:concl}

We have constructed a sample of stars based on the cross match of the
recent first Gaia data release known as TGAS to the RAVE spectroscopic survey, and identified
a subset as halo stars based on their RAVE metallicity. We analysed
their kinematics and demonstrated with a velocity correlation function, the existence of
a significant excess of pairs of stars with small velocity differences
compared to the number expected for random distributions obtained by
re-shuffling the various velocity components of the stars in our sample. The statistical significance of the
signal is 3.7$\sigma$ for separations smaller than 20~km/s, and
8.8$\sigma$ for 40 km/s separations.

We then determined the distribution of our sample of halo stars in
``Integrals of Motion'' space, defined by two components of the angular
momentum, $L_\perp$ and $L_z$ and by the energy of the stars, which
was computed for a reasonable guess of the Galactic potential. The
distribution of stars in this space is complex and shows a high degree
of structure. Firstly, we established the presence of a dominant
retrograde component for stars somewhat less bound than the Sun. The
probability that $\sim$ 60\% or more of the stars with such orbital
characteristics occurs by chance as estimated from the randomised
smooth sets is smaller than 1/1000. For the more bound halo stars, we
identify at least 10 statistically significant overdensities, whose
probability derived using the randomised sets, is smaller than 1\%. Of
these overdensities, the most prominent appears to be associated to
the metal-poor tail of the disk(s).

We have also performed comparisons to cosmological simulations. The
level of substructure revealed by the velocity correlation function for our sample
is comparable to that found in solar neighbourhood like volumes with
similar numbers of stars in the stellar halos of the Aquarius
simulations by \citet{Lowing2015}. This indicates that it is plausible
that all of the Galactic stellar halo was built via accretion, as this
is the only channel considered in these simulations. We have also
established the frequency of occurrence of retrograde outer halos with
similar predominance as estimated from our halo sample. In the
Illustris simulations \citep{Vogelsberger2014a}, less than 1\% of the
Milky Way mass galaxies have outer halos where more than $\sim 60\%$
of the stars have retrograde motions. At face value, this is very
intriguing. However, this comparison suffers from several issues, such
as incompleteness in the observational sample and numerical resolution
limitations of the simulations. Nonetheless, the predominance of the
retrograde halo is striking in itself, and has been found perhaps less
dramatically by several studies independently, indicating that it
is likely a major component of the Galactic halo.

This first analysis of data obtained by the Gaia satellite mission has
thus revealed many exciting results. We look forward to better
understanding what our findings imply, from the dynamical and chemical
perspectives, for the history of the Milky Way. There is plenty of work
to do before the second Gaia data release becomes available.

\begin{acknowledgements}
  It is a pleasure to thank Lorenzo Posti and Anthony Brown for the
  numerous enlightening discussions. We are grateful to the referee
  for a prompt and constructive report. AH acknowledges
  financial support from a VICI grant from the Netherlands
  Organisation for Scientific Research, NWO. MB, JV and AH are
  grateful to NOVA for financial support. This work has made use of
  data from the European Space Agency (ESA) mission {\it Gaia}
  (\url{http://www.cosmos.esa.int/gaia}), processed by the {\it Gaia}
  Data Processing and Analysis Consortium (DPAC,
  \url{http://www.cosmos.esa.int/web/gaia/dpac/consortium}). Funding
  for the DPAC has been provided by national institutions, in
  particular the institutions participating in the {\it Gaia}
  Multilateral Agreement.

\end{acknowledgements}

%
%

\bibliographystyle{aa} 


\begin{appendix}

\section{Estimating the bias on the distance obtained by inverting 
the trigonometric parallax}
\label{App:distance-bias}

 In the ideal case of error-free measurements, the distance
  to a star is simply the inverse of its measured trigonometric
  parallax. In reality however, measurements do have errors, and this
  makes the calculation of the distance significantly more
  involved. There has been some debate in the literature on what is
  the best way to obtain reliable distances to stars using their
  parallaxes, and what biases one can expect when inverting the
  parallax to obtain a distance
  \citep[e.g.][]{Arenou99,Smith06,Astraatmadja16}.

\begin{figure}
\centering
\includegraphics[width=8.5cm]{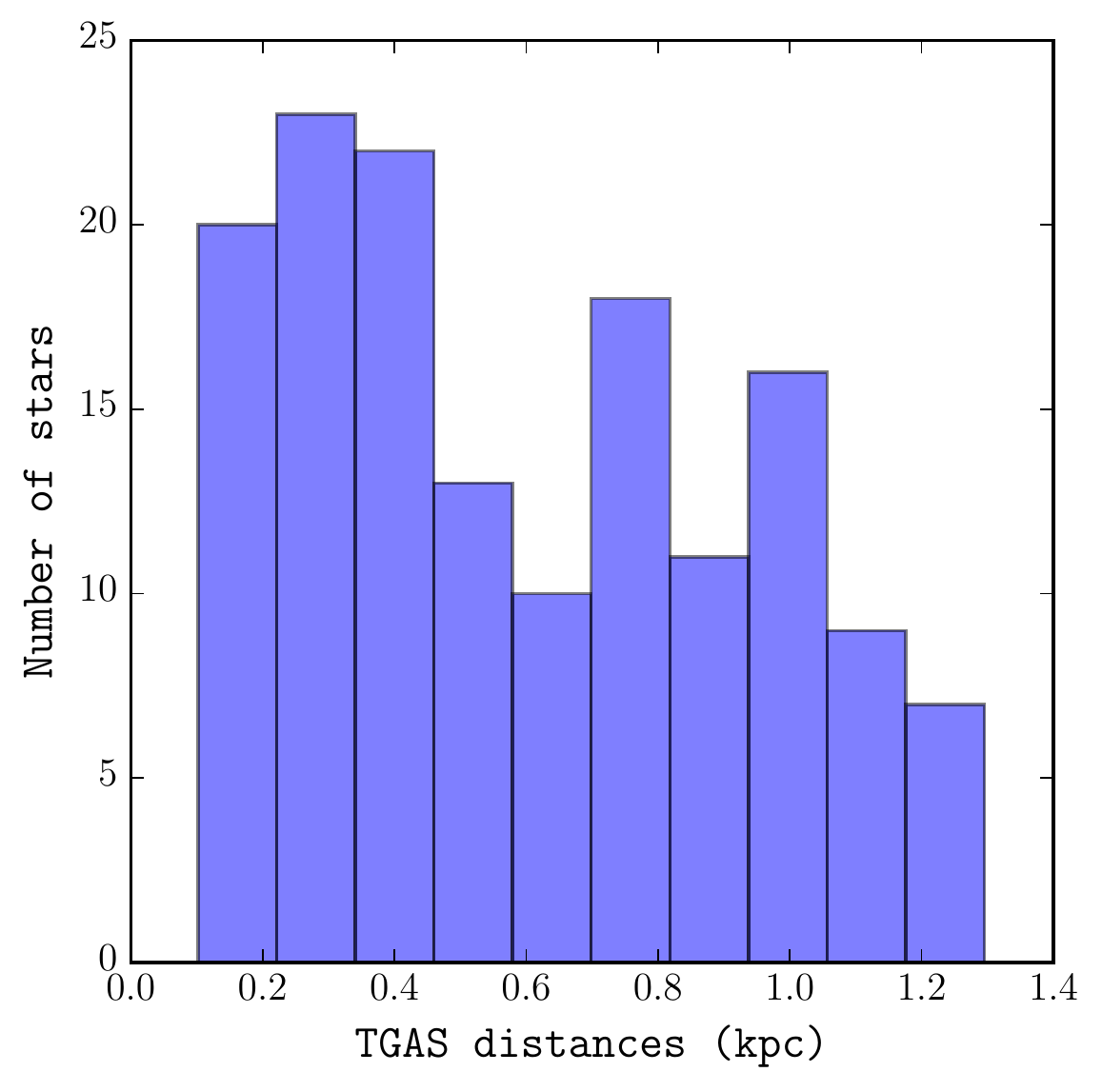} 
\caption{Distribution of distances obtained by inverting the trigonometric parallaxes for our sample of 
halo stars and for which the TGAS parallax is more precise than the RAVE parallax estimate.}
\label{fig:dist_tgas}
\end{figure}

  For the purpose of the analysis presented in this paper, we use
  distances obtained by inverting the parallaxes both from RAVE and
  from the TGAS datasets. For the RAVE stars, as discussed earlier,
  \citet{Binney2014} has shown that the best distance estimate is
  obtained by inverting the parallax estimated by the RAVE
  pipeline. Our focus in this Appendix is therefore on those halo
  stars in our sample for which the TGAS parallax is more precise than
  the RAVE parallax. These stars are all located within
  1.2 kpc from the Sun, as shown in Fig.~\ref{fig:dist_tgas}. 

  In order to quantify the bias we may introduce by inverting the TGAS 
  parallaxes, we have performed the following test. Using the \Gaia
  Universe Model Snapshot \citep[GUMS,][]{Robin2012GUMS} we select all
  halo stars in the model within 3~kpc from the Sun that are within
  the TGAS magnitude limits. We consider a larger distance
  range to explore more broadly the issue of inverting trigonometric
  parallaxes. GUMS here represents a perfect model, and all stellar
  quantities available are error-free, meaning one can convert from
  distance to parallax by taking its inverse. We then convolve these
  true parallaxes with the errors of the halo stars in the sample we
  defined in Section~\ref{sec:halo_sel}, and assuming the errors are
  Gaussian. The convolution is repeated 1000 times, so we have 1000
  different realisations of the same GUMS stars. Finally we calculated
  the ``measured'' distances by inverting the ``individual measurements'' of the parallaxes.

\begin{figure*}
\centering
\includegraphics[width=16cm]{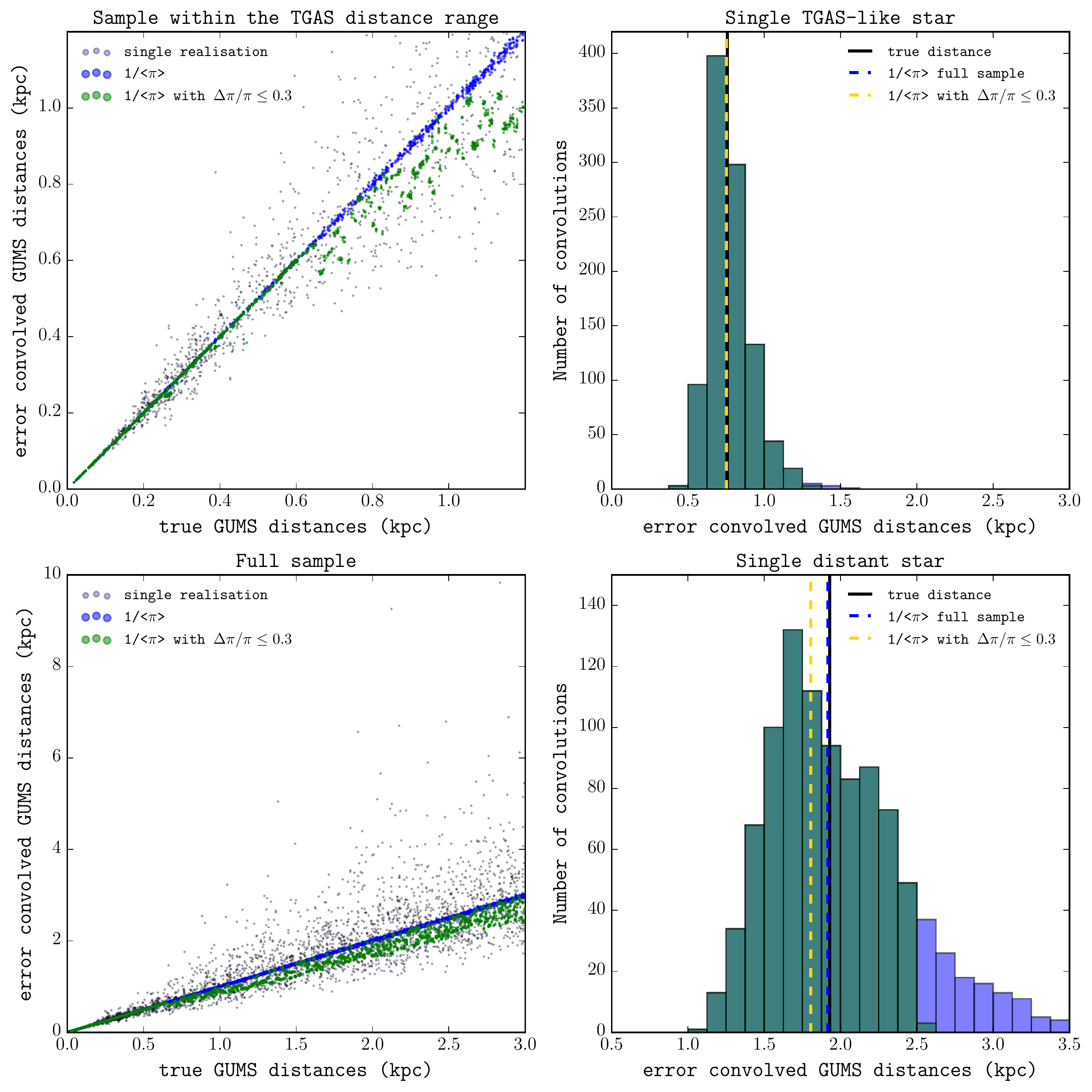} 
\caption{Left panels: the true distances vs a (random) set of
  error-convolved distances for a sample of halo stars from the \Gaia
  Universe Model Snapshot. The top panels cover the distance range
  for the stars in our halo sample with parallaxes from TGAS, while
  the bottom panels does so for the full halo sample. The blue points
  in these panels represent the inverted average value of the
  parallaxes obtained from all 1000 convolved sets for each star, and
  they lie nearly perfectly on the 1:1 line, while the green points are
  the averages obtained using only those realisations for which the
  relative parallax error is smaller than 30\%. Right panels: The
  distribution of distances from the 1000 error convolved parallaxes
  for a single star, at a distance of 0.75 kpc representative of a
  TGAS star (top), and at a distance of 1.9~kpc for a more distant
  object (bottom). One can see that the true distance to the star
  (solid vertical line), matches extremely well the distance obtained
  by inverting the parallax (dashed vertical line), and this in fact
  coincides well with the mean of the distance distribution.}
\label{fig:distance-bias}
\end{figure*}

Figure \ref{fig:distance-bias} shows that there is virtually no
difference between the true distance for the stars in the GUMS
dataset, and the mean distance obtained by inverting the ``mean''
parallax of all 1000 error convolved sets.  In fact, the average
absolute difference between these two quantities is smaller than
10~pc. The left panels of Fig.~\ref{fig:distance-bias} shows the true
distances of GUMS stars versus one set of the error convolved
distances, as black dots. The top panel corresponds to the distance
range probed by the TGAS stars in our sample, while the bottom panel
is for distances upto 3~kpc. In both panels, the blue points represent the mean
values obtained by inverting the mean parallax of all 1000 convolved
sets, and these lie perfectly on the 1:1 relation. The green points
correspond to the mean obtained when applying a 30\% cut on the
relative parallax error for each realisation.

The bottom left panel on Fig.~\ref{fig:distance-bias} especially shows
that the black dots (which represent distances obtained from a single
realisation of the error convolved parallaxes) are not evenly
scattered around the 1:1 line. This tells us that, when convolving
true parallaxes with errors and then inverting them to get the
distance, the most likely value obtained for the distance often
underestimates the true distance. On the other hand, the distance can
be scattered more towards larger values than it is towards
smaller. 

This effect is clearly shown in the right handside panels of
Fig.~\ref{fig:distance-bias}.  The distance distribution of a single
star for all 1000 error convolved sets is shown in the top for a TGAS
star at a true distance of 0.75 kpc, and for a star farther away at
1.9~kpc in the bottom panel. In both cases, the solid and dashed
vertical lines mark the true distance and distance obtained by inverting the 
mean of the parallax distribution respectively, of the considered star, 
without (blue) and with (yellow) a relative parallax error cut. Since the TGAS 
halo stars in our sample are located closer than $\sim 1.2$ kpc, we can conclude
that the effect is negligible.

More generally, if we can assume that the errors on the parallax are
Gaussian, then by inverting the mean parallax to obtain a distance, it
is possible to recover the true distance well. In the case of Gaussian
errors, the mean and the maximum likelihood estimator of the parallax
coincide. Note that it would not seem to be wise under these
circumstances, to attempt to derive a distance estimate from the
inversion of each of the individual trigonometric parallaxes, since
the maximum likelihood estimator of the distances obtained is
different from the mean of their distribution. Since the
parallax error distribution has not been characterised yet for the
TGAS sample, we are forced to make the simple assumption of Gaussian
errors.  Future data releases will allow us to establish more robustly
if there are biases that need to be considered.

\newpage

\section{Stars members of the newly identified substructures}
\label{App:starlist}

\begin{table}[h]
\centering
\caption{Positions and Tycho ids for the stars belonging to a substructure
identified as the disk Fig.~\ref{fig:watershed}. A total of 30 stars are marked
as disk members.}
\label{tab:starlist-disk}
\begin{tabular}{llll}
\hline
\hline
Substructure & l     & b     & Tycho-id             \\
             & (deg) & (deg) & \\
\hline
Disk & 343.07 & -80.73 & 6993-953-1 \\
Disk & 78.83  & -77.83 & 5842-477-1 \\
Disk & 49.14  & -68.26 & 6399-1131-1 \\
Disk & 149.21 & -62.99 & 4685-1855-1 \\
Disk & 57.53  & -53.18 & 5811-1294-1 \\
Disk & 227.38 & -27.39 & 6476-533-1 \\
Disk & 304.02 & 38.43  & 6701-78-1 \\
Disk & 281.97 & 47.61  & 5523-465-1 \\
Disk & 322.26 & 56.87  & 4969-862-1 \\
Disk & 294.60 & -33.41 & 9371-521-1 \\
Disk & 263.87 & -50.31 & 8066-340-1 \\
Disk & 266.20 & -38.08 & 8513-362-1 \\
Disk & 223.27 & -45.93 & 6459-1312-1 \\
Disk & 282.84 & -14.05 & 8939-1294-1 \\
Disk & 282.77 & -5.61  & 8942-2009-1 \\
Disk & 281.54 & -12.67 & 8935-146-1 \\
Disk & 249.72 & -8.70  & 7117-324-1 \\
Disk & 246.55 & -12.36 & 7098-1232-1 \\
Disk & 243.54 & -9.18  & 7103-1352-1 \\
Disk & 238.01 & -8.52  & 6527-97-1 \\
Disk & 237.09 & -6.15  & 6524-1307-1 \\
Disk & 238.24 & 6.54   & 6003-1712-1 \\
Disk & 243.20 & 17.77  & 6013-995-1 \\
Disk & 294.52 & 21.16  & 7759-888-1 \\
Disk & 348.00 & 31.78  & 6175-369-1 \\
Disk & 325.96 & 30.95  & 6739-220-1 \\
Disk & 307.95 & -26.20 & 9527-1034-1 \\
Disk & 325.36 & -62.64 & 8463-440-1 \\
Disk & 26.27  & -65.29 & 6974-198-1 \\
Disk & 35.35  & -38.44 & 5784-820-1 \\
\hline
\end{tabular}
\end{table}

\begin{table}
\centering
\caption{Positions and Tycho ids for the stars belonging to VelHel-1.
This structure comprises 28 stars.}
\label{tab:starlist-1}
\begin{tabular}{llll}
\hline
\hline
Substructure & l     & b     & Tycho-id             \\
             & (deg) & (deg) & \\
\hline
VelHel-1 & 42.36  & -66.08 & 6401-458-1 \\
VelHel-1 & 230.62 & -24.28 & 6494-492-1 \\
VelHel-1 & 236.93 & -13.73 & 6529-97-1 \\
VelHel-1 & 207.78 & -41.22 & 5318-1280-1 \\
VelHel-1 & 304.49 & 40.57  & 6117-109-1 \\
VelHel-1 & 308.03 & 40.61  & 6118-114-1 \\
VelHel-1 & 272.59 & 36.25  & 6083-431-1 \\
VelHel-1 & 271.69 & 52.69  & 4936-193-1 \\
VelHel-1 & 285.19 & -45.49 & 8866-217-1 \\
VelHel-1 & 305.02 & -53.06 & 8844-799-1 \\
VelHel-1 & 258.47 & -37.73 & 8084-946-1 \\
VelHel-1 & 219.82 & -40.47 & 5897-740-1 \\
VelHel-1 & 290.66 & -73.20 & 7540-650-1 \\
VelHel-1 & 175.86 & -77.54 & 5854-672-1 \\
VelHel-1 & 285.99 & 12.83  & 8200-1944-1 \\
VelHel-1 & 240.81 & -12.22 & 7089-375-1 \\
VelHel-1 & 237.93 & 11.45  & 5997-1706-1 \\
VelHel-1 & 333.81 & 28.22  & 6755-1118-1 \\
VelHel-1 & 346.42 & 38.86  & 5588-1160-1 \\
VelHel-1 & 325.61 & -30.75 & 9310-1481-1 \\
VelHel-1 & 329.60 & -34.30 & 9099-1107-1 \\
VelHel-1 & 333.82 & -35.27 & 9091-881-1 \\
VelHel-1 & 330.15 & -38.03 & 9109-92-1 \\
VelHel-1 & 344.70 & -66.77 & 8019-29-1 \\
VelHel-1 & 352.87 & -44.93 & 8423-827-1 \\
VelHel-1 & 34.00  & -61.26 & 6964-271-1 \\
VelHel-1 & 346.74 & -11.92 & 8357-2214-1 \\
VelHel-1 & 30.49  & -47.97 & 6375-730-1 \\
\hline
\end{tabular}
\end{table}

\begin{table}
\centering
\caption{Positions and Tycho ids for the stars belonging to VelHel-2.
This structure comprises 4 stars.}
\label{tab:starlist-2}
\begin{tabular}{llll}
\hline
\hline
Substructure & l     & b     & Tycho-id             \\
             & (deg) & (deg) & \\
\hline
VelHel-2 & 301.17 & -38.73 & 9351-1553-1 \\
VelHel-2 & 250.96 & -20.12 & 7622-1514-1 \\
VelHel-2 & 348.31 & 28.93  & 6193-1514-1 \\
VelHel-2 & 341.43 & -33.02 & 8784-1728-1 \\
\hline
\end{tabular}
\end{table}

\begin{table}
\centering
\caption{Positions and Tycho ids for the stars belonging to the substructures
VelHel-3. A total of 8 stars form this structure.}
\label{tab:starlist-3}
\begin{tabular}{llll}
\hline
\hline
Substructure & l     & b     & Tycho-id             \\
             & (deg) & (deg) & \\
\hline
VelHel-3 & 284.75 & -55.98 & 8489-732-1 \\
VelHel-3 & 272.78 & -58.89 & 8056-192-1 \\
VelHel-3 & 304.40 & -58.33 & 8472-779-1 \\
VelHel-3 & 305.22 & -67.55 & 8034-464-1 \\
VelHel-3 & 238.67 & -57.46 & 7026-864-1 \\
VelHel-3 & 332.08 & -51.06 & 8826-198-1 \\
VelHel-3 & 4.81   & -37.02 & 7963-155-1 \\
VelHel-3 & 23.22  & -29.13 & 6340-429-1 \\
\hline
\end{tabular}
\end{table}

\begin{table}
\centering
\caption{Positions and Tycho ids for the stars belonging to VelHel-4.
This structure consists of 7 stars.}
\label{tab:starlist-4}
\begin{tabular}{llll}
\hline
\hline
Substructure & l     & b     & Tycho-id             \\
             & (deg) & (deg) & \\
\hline
VelHel-4 & 4.15   & 27.29  & 5617-720-1 \\
VelHel-4 & 287.63 & 6.33   & 8616-807-1 \\
VelHel-4 & 292.62 & 12.17  & 8224-1458-1 \\
VelHel-4 & 260.62 & 18.06  & 6614-789-1 \\
VelHel-4 & 299.16 & 13.27  & 8239-728-1 \\
VelHel-4 & 346.49 & -39.76 & 8420-259-1 \\
VelHel-4 & 6.61   & -28.39 & 7448-744-1 \\
\end{tabular}
\end{table}

\begin{table}
\centering
\caption{Positions and Tycho ids for the stars belonging to VelHel-5.
This structure comprises 17 stars.}
\label{tab:starlist-5}
\begin{tabular}{llll}
\hline
\hline
Substructure & l     & b     & Tycho-id             \\
             & (deg) & (deg) & \\
\hline
VelHel-5 & 142.56 & -87.86 & 6422-422-1 \\
VelHel-5 & 27.16  & -68.66 & 6976-6-1 \\
VelHel-5 & 279.42 & -30.02 & 9163-387-1 \\
VelHel-5 & 250.21 & -41.13 & 7592-1104-1 \\
VelHel-5 & 242.45 & -38.51 & 7587-1369-1 \\
VelHel-5 & 235.60 & -53.41 & 7031-616-1 \\
VelHel-5 & 219.65 & -61.17 & 6441-249-1 \\
VelHel-5 & 275.39 & 16.66  & 7710-2266-1 \\
VelHel-5 & 241.14 & -15.41 & 7087-114-1 \\
VelHel-5 & 315.92 & -15.33 & 9281-1877-1 \\
VelHel-5 & 354.84 & 28.58  & 6187-147-1 \\
VelHel-5 & 340.84 & -41.30 & 8797-888-1 \\
VelHel-5 & 320.17 & -52.10 & 9126-444-1 \\
VelHel-5 & 320.41 & -52.52 & 9126-1178-1 \\
VelHel-5 & 346.27 & -67.09 & 8016-762-1 \\
VelHel-5 & 35.40  & -59.26 & 6392-48-1 \\
VelHel-5 & 351.68 & -13.14 & 7909-380-1 \\
\hline
\end{tabular}
\end{table}

\begin{table}
\centering
\caption{Positions and Tycho ids for the stars belonging to VelHel-6.
A total of 69 are marked to belong to this structure.}
\label{tab:starlist-6}
\begin{tabular}{llll}
\hline
\hline
Substructure & l     & b     & Tycho-id             \\
             & (deg) & (deg) & \\
\hline
VelHel-6 & 19.71  & -82.75 & 6419-200-1 \\
VelHel-6 & 20.25  & -73.97 & 6987-302-1 \\
VelHel-6 & 38.16  & -80.37 & 6415-1008-1 \\
VelHel-6 & 30.63  & -70.59 & 6976-513-1 \\
VelHel-6 & 80.29  & -65.98 & 5830-840-1 \\
VelHel-6 & 63.60  & -56.31 & 5241-786-1 \\
VelHel-6 & 277.68 & 34.27  & 6650-1108-1 \\
VelHel-6 & 268.97 & 33.59  & 6082-404-1 \\
VelHel-6 & 267.33 & 35.88  & 6079-657-1 \\
VelHel-6 & 280.14 & 45.88  & 5523-1415-1 \\
VelHel-6 & 252.68 & 46.82  & 4916-435-1 \\
VelHel-6 & 8.91   & 31.55  & 5042-508-1 \\
VelHel-6 & 297.24 & -30.20 & 9492-2130-1 \\
VelHel-6 & 294.77 & -34.75 & 9371-1098-1 \\
VelHel-6 & 287.70 & -39.89 & 9155-38-1 \\
VelHel-6 & 292.53 & -41.59 & 9151-604-1 \\
VelHel-6 & 292.86 & -42.97 & 9147-26-1 \\
VelHel-6 & 273.20 & -46.58 & 8506-763-1 \\
VelHel-6 & 270.67 & -45.18 & 8507-1765-1 \\
VelHel-6 & 274.36 & -59.92 & 8055-583-1 \\
VelHel-6 & 269.36 & -55.20 & 8491-12-1 \\
VelHel-6 & 263.41 & -29.55 & 8524-318-1 \\
VelHel-6 & 247.09 & -33.61 & 7603-683-1 \\
VelHel-6 & 245.65 & -49.91 & 7573-1124-1 \\
VelHel-6 & 220.38 & -43.39 & 6457-2149-1 \\
VelHel-6 & 293.69 & -62.75 & 8474-959-1 \\
VelHel-6 & 313.24 & -62.84 & 8465-1053-1 \\
VelHel-6 & 240.66 & -69.07 & 7009-568-1 \\
VelHel-6 & 282.53 & -74.36 & 7544-778-1 \\
VelHel-6 & 237.61 & -60.41 & 7025-394-1 \\
VelHel-6 & 242.39 & -62.72 & 7018-102-1 \\
VelHel-6 & 210.95 & -71.25 & 6433-2034-1 \\
VelHel-6 & 188.00 & -71.81 & 5856-811-1 \\
VelHel-6 & 185.26 & -60.88 & 5288-567-1 \\
VelHel-6 & 286.42 & -8.02  & 8951-822-1 \\
VelHel-6 & 270.22 & -16.59 & 8560-1461-1 \\
VelHel-6 & 287.67 & 12.42  & 8217-1411-1 \\
VelHel-6 & 277.01 & 13.77  & 7718-2873-1 \\
VelHel-6 & 275.47 & 17.80  & 7710-9-1 \\
VelHel-6 & 249.71 & -12.39 & 7630-809-1 \\
VelHel-6 & 301.23 & 15.57  & 8236-795-1 \\
VelHel-6 & 294.09 & 21.11  & 7759-1254-1 \\
VelHel-6 & 303.06 & 26.72  & 7260-938-1 \\
VelHel-6 & 326.33 & 26.03  & 7299-914-1 \\
\hline
\end{tabular}
\end{table}

\addtocounter{table}{-1}
\begin{table}
\centering
\caption{Table~\ref{tab:starlist-6} continued.}
\begin{tabular}{llll}
\hline
\hline
Substructure & l     & b     & Tycho-id             \\
             & (deg) & (deg) & \\
\hline
VelHel-6 & 329.10 & 28.30  & 6757-705-1 \\
VelHel-6 & 347.44 & 30.58  & 6188-1320-1 \\
VelHel-6 & 344.24 & 36.83  & 6169-218-1 \\
VelHel-6 & 316.62 & -29.26 & 9463-644-1 \\
VelHel-6 & 322.68 & -32.09 & 9315-572-1 \\
VelHel-6 & 312.43 & -44.84 & 9342-529-1 \\
VelHel-6 & 312.99 & -49.79 & 9133-1330-1 \\
VelHel-6 & 344.42 & -43.53 & 8806-147-1 \\
VelHel-6 & 345.80 & -40.70 & 8433-1057-1 \\
VelHel-6 & 325.15 & -52.01 & 9125-625-1 \\
VelHel-6 & 348.27 & -58.37 & 8446-381-1 \\
VelHel-6 & 351.94 & -68.92 & 8013-15-1 \\
VelHel-6 & 349.48 & -52.13 & 8438-853-1 \\
VelHel-6 & 11.23  & -65.38 & 7509-805-1 \\
VelHel-6 & 344.22 & -30.05 & 8775-1063-1 \\
VelHel-6 & 344.05 & -31.57 & 8776-1875-1 \\
VelHel-6 & 348.05 & -30.74 & 8399-421-1 \\
VelHel-6 & 5.27   & -37.14 & 7468-171-1 \\
VelHel-6 & 357.91 & -30.31 & 7955-1803-1 \\
VelHel-6 & 10.42  & -31.68 & 7446-1142-1 \\
VelHel-6 & 355.32 & -20.58 & 7925-894-1 \\
VelHel-6 & 16.53  & -46.92 & 7474-266-1 \\
VelHel-6 & 39.47  & -43.01 & 5799-506-1 \\
VelHel-6 & 17.35  & -30.75 & 6911-525-1 \\
VelHel-6 & 19.15  & -31.64 & 6911-244-1 \\
\hline
\end{tabular}
\end{table}

\begin{table}
\centering
\caption{Positions and Tycho ids for the stars belonging to VelHel-7.
There are 43 stars in this structure.}
\label{tab:starlist-7}
\begin{tabular}{llll}
\hline
\hline
Substructure & l     & b     & Tycho-id             \\
             & (deg) & (deg) & \\
\hline
VelHel-7 & 52.98  & -44.25 & 5230-635-1 \\
VelHel-7 & 289.98 & 29.40  & 7222-638-1 \\
VelHel-7 & 282.35 & 31.79  & 6659-325-1 \\
VelHel-7 & 289.11 & -36.23 & 9364-1205-1 \\
VelHel-7 & 280.32 & -32.01 & 9166-247-1 \\
VelHel-7 & 296.26 & -51.57 & 8855-1243-1 \\
VelHel-7 & 293.96 & -53.28 & 8853-1169-1 \\
VelHel-7 & 278.32 & -50.21 & 8859-145-1 \\
VelHel-7 & 270.79 & -57.15 & 8056-155-1 \\
VelHel-7 & 237.24 & -40.87 & 7045-1116-1 \\
VelHel-7 & 287.98 & -14.75 & 9204-1228-1 \\
VelHel-7 & 272.50 & -14.42 & 8578-2046-1 \\
VelHel-7 & 290.44 & 8.96   & 8226-106-1 \\
VelHel-7 & 276.58 & 8.22   & 8183-1066-1 \\
VelHel-7 & 268.01 & 18.50  & 7174-1341-1 \\
VelHel-7 & 311.02 & -17.01 & 9433-453-1 \\
VelHel-7 & 312.40 & -15.95 & 9429-908-1 \\
VelHel-7 & 304.96 & -12.54 & 9426-763-1 \\
VelHel-7 & 328.09 & -12.73 & 9043-1518-1 \\
VelHel-7 & 322.46 & -11.36 & 9049-2020-1 \\
VelHel-7 & 333.71 & -12.27 & 8736-17-1 \\
VelHel-7 & 308.45 & 8.82   & 8662-1638-1 \\
VelHel-7 & 320.55 & 22.87  & 7293-1952-1 \\
VelHel-7 & 335.21 & 29.77  & 6751-684-1 \\
VelHel-7 & 345.26 & 30.78  & 6179-259-1 \\
VelHel-7 & 351.24 & 28.87  & 6190-766-1 \\
VelHel-7 & 347.01 & 41.97  & 5583-971-1 \\
VelHel-7 & 354.80 & 48.24  & 4987-339-1 \\
VelHel-7 & 307.64 & -37.13 & 9485-48-1 \\
VelHel-7 & 313.22 & -28.44 & 9471-893-1 \\
VelHel-7 & 312.24 & -47.15 & 9339-1999-1 \\
VelHel-7 & 326.00 & -45.80 & 9120-909-1 \\
VelHel-7 & 332.87 & -35.93 & 9104-985-1 \\
VelHel-7 & 328.03 & -26.55 & 9292-421-1 \\
VelHel-7 & 333.61 & -52.50 & 8826-1265-1 \\
VelHel-7 & 336.63 & -69.58 & 8021-220-1 \\
VelHel-7 & 10.83  & -59.37 & 7500-619-1 \\
VelHel-7 & 1.28   & -31.47 & 7952-937-1 \\
VelHel-7 & 42.07  & -48.70 & 5808-987-1 \\
VelHel-7 & 29.47  & -25.09 & 5752-416-1 \\
VelHel-7 & 32.78  & -27.31 & 5762-477-1 \\
VelHel-7 & 354.42 & 27.85  & 6187-55-1 \\
VelHel-7 & 319.77 & 42.50  & 6126-688-1 \\
\hline
\end{tabular}
\end{table}

\begin{table}
\centering
\caption{Positions and Tycho ids for the stars belonging to VelHel-8. This
structure contains 8 stars.}
\label{tab:starlist-8-9}
\begin{tabular}{llll}
\hline
\hline
Substructure & l     & b     & Tycho-id             \\
             & (deg) & (deg) & \\
\hline
VelHel-8 & 234.61 & -17.56 & 6514-2428-1 \\
VelHel-8 & 224.65 & -33.92 & 6473-1182-1 \\
VelHel-8 & 38.68  & -23.79 & 5175-874-1 \\
VelHel-8 & 287.03 & 12.23  & 8217-1855-1 \\
VelHel-8 & 250.58 & -17.49 & 7632-139-1 \\
VelHel-8 & 345.28 & 27.19  & 6196-462-1 \\
VelHel-8 & 7.09   & -59.28 & 7503-692-1 \\
VelHel-8 & 24.85  & -27.98 & 6336-20-1 \\
\hline
\end{tabular}
\end{table}

\begin{table}
\centering
\caption{Positions and Tycho ids for the stars belonging to VelHel-9.
This structure comprises 5 stars.}
\label{tab:starlist-1}
\begin{tabular}{llll}
\hline
\hline
Substructure & l     & b     & Tycho-id             \\
             & (deg) & (deg) & \\
\hline
VelHel-9 & 277.57 & 36.63  & 6091-633-1 \\
VelHel-9 & 232.84 & -54.85 & 7024-1104-1 \\
VelHel-9 & 241.85 & 20.22  & 5456-610-1 \\
VelHel-9 & 317.55 & -41.09 & 9332-352-1 \\
VelHel-9 & 336.69 & -26.69 & 8772-745-1 \\
\hline
\end{tabular}
\end{table}

\end{appendix}

\end{document}